\documentclass[twocolumn]{aastex63}

\usepackage{mathtools}

\newcommand{\hi}{H\,{\sc i}}
\newcommand\kms{km$\,$s$^{-1}$}

\usepackage[multiple]{footmisc}

\received{}
\revised{}
\accepted{}
\shorttitle{Faint, Quenched Dwarfs at D$\approx$2 Mpc}
\shortauthors{Sand et al.}

\graphicspath{{./}{figures/}}

\begin{document}

\title{Three Quenched, Faint Dwarf Galaxies in the Direction of NGC~300: New Probes of Reionization and Internal Feedback
}

\correspondingauthor{D. J. Sand}
\email{dsand@arizona.edu}

\newcommand{\Thacher}{\affiliation{Thacher Observatory, Thacher School, 5025 Thacher Rd. Ojai, CA 93023, USA}}
\newcommand{\UA}{\affiliation{Steward Observatory, University of Arizona, 933 North Cherry Avenue, Tucson, AZ 85721-0065, USA}}
\newcommand{\NU}{\affiliation{Center for Interdisciplinary Exploration and Research in Astrophysics and Department of Physics and Astronomy, \\ Northwestern University, 2145 Sheridan Road, Evanston, IL 60208-3112, USA}}
\newcommand{\UCDavis}{\affiliation{Department of Physics, University of California, 1 Shields Avenue, Davis, CA 95616-5270, USA}}
\newcommand{\Padova}{\affiliation{Department of Physics and Astronomy Galileo Galilei, University of Padova, Vicolo dell'Osservatorio, 3, I-35122 Padova, Italy}}
\newcommand{\INAF}{\affiliation{INAF Osservatorio Astronomico di Padova, Vicolo dell'Osservatorio 5, I-35122 Padova, Italy}}
\newcommand{\INAFbol}{\affiliation{INAF - Osservatorio di Astrofisica e Scienza dello Spazio - Via Piero Gobetti 93/3, I-40129 Bologna, Italy}}
\newcommand{\LPL}{\affiliation{Lunar and Planetary Lab, Department of Planetary Sciences, University of Arizona, Tucson, AZ 85721, USA}}
\newcommand{\NOAO}{\affiliation{National Optical Astronomy Observatory, 950 North Cherry Avenue, Tucson, AZ 85719, USA}}
\newcommand{\PATAU}{\affiliation{The School of Physics and Astronomy, Tel Aviv University, Tel Aviv 69978, Israel}}
\newcommand{\TTU}{\affiliation{Department of Physics and Astronomy, Texas Tech University, Box 1051, Lubbock, TX 79409-1051, USA}}
\newcommand{\OSU}{\affiliation{Department  of  Astronomy,  The  Ohio  State University,  140  W.  18th  Ave.,  Columbus,  OH43210, USA}}
\newcommand{\LBT}{\affiliation{Large Binocular Telescope Observatory, 933 North Cherry Avenue, Tucson, AZ, USA}}
\newcommand{\RMC}{\affiliation{Department of Physics and Space Science Royal Military College of Canada P.O. Box 17000, Station Forces Kingston, ON K7K 7B4, Canada}}
\newcommand{\ASU}{\affiliation{School of Earth and Space Exploration, Arizona State University, Tempe, AZ 85287, USA}}
\newcommand{\MMT}{\affiliation{MMT Observatory, PO Box 210065, University of Arizona, Tucson, AZ 85721-0065, USA}}
\newcommand{\NAU}{\affiliation{Department of Physics and Astronomy, Northern Arizona University, P.O. Box 6010, Flagstaff, AZ 86011, USA}}
\newcommand{\UAOptSci}{\affiliation{College of Optical Sciences, University of Arizona, 1630 E University Blvd, Tucson, AZ 85719, USA}}
\newcommand{\UNC}{\affiliation{Department of Physics and Astronomy, University of North Carolina at Chapel Hill, Chapel Hill, NC 27599, USA}}
\newcommand{\MSU}{\affiliation{Center for Data Intensive and Time Domain Astronomy, Department  of  Physics  and  Astronomy,  Michigan  State  University,East Lansing, MI 48824, USA}}
\newcommand{\UCSC}{\affiliation{Department of Astronomy and Astrophysics, University of California, Santa Cruz, CA 95064, USA}}
\newcommand{\STScI}{\affiliation{Space Telescope Science Institute, 3700 San Martin Drive, Baltimore, MD 21218, USA}}
\newcommand{\Brandeis}{\affiliation{Department of Physics, Brandeis University, Waltham, MA 02453, USA}}
\newcommand{\LCO}{\affiliation{Las Cumbres Observatory, 6740 Cortona Drive, Suite 102, Goleta, CA 93117-5575, USA}}
\newcommand{\UToronto}{\affiliation{Department of Astronomy and Astrophysics, University of Toronto, 50 St. George Street, Toronto, Ontario, M5S 3H4 Canada}}
\newcommand{\NotreDame}{\affiliation{Department of Physics, University of Notre Dame, Notre Dame, IN 46556, USA}}
\newcommand{\UMN}{\affiliation{College of Science \& Engineering, Minnesota Institute for Astrophysics, University of Minnesota, 115 Union St. SE, Minneapolis, MN 55455, USA}}
\newcommand{\UT}{\affiliation{Department of Astronomy, University of Texas at Austin, Austin, TX 78712, USA}}
\newcommand{\JHU}{\affiliation{The Johns Hopkins University, Baltimore, MD 21218, USA}}
\newcommand{\VAT}{\affiliation{Vatican Observatory, 00120 Citt\`{a} del Vaticano, Vatican City State  }}
\newcommand{\HF}{\affiliation{Hubble Fellow}}
\newcommand{\Carnegie}{\affiliation{The Observatories of the Carnegie Institution for Science, 813 Santa Barbara St., Pasadena, CA 91101, USA}}

\author[0000-0003-4102-380X]{David~J. Sand}
\UA

\author[0000-0001-9649-4815]{\textsc{Bur\c{c}{\rlap{\.}\i}n Mutlu-Pakd{\rlap{\.}\i}l}}
\affil{Department of Physics and Astronomy, Dartmouth College, 6127 Wilder Laboratory, Hanover, NH 03755, USA}

\author[0000-0002-5434-4904]{Michael G. Jones}
\UA
%\affiliation{Steward Observatory, University of Arizona, 933 North Cherry Avenue, Rm. N204, Tucson, AZ 85721-0065, USA}

\author[0000-0001-8855-3635]{Ananthan Karunakaran}
\affiliation{Department of Astronomy \& Astrophysics, University of Toronto, Toronto, ON M5S 3H4, Canada}

\author[0000-0003-0123-0062]{Jennifer E. Andrews}
\affiliation{International Gemini Observatory/NSF NOIRLab, 670 N. A'ohoku Place, Hilo, Hawai'i, 96720, USA}

\author[0000-0001-8354-7279]{Paul Bennet}
\affiliation{Space Telescope Science Institute, 3700 San Martin Drive, Baltimore, MD 21218, USA}

\author[0000-0002-1763-4128]{Denija Crnojevi\'{c}}
\affil{Department of Physics \& Astronomy, University of Tampa, 401 West Kennedy Boulevard, Tampa, FL 33606, USA}

\author[0000-0003-2536-5092]{Giuseppe Donatiello}
\affil{Unione Astrofili Italiani P.I. Sezione Nazionale di Ricerca Profondo Cielo, 72024 Oria, Italy}

\author[0000-0001-8251-933X]{Alex Drlica-Wagner}
\affil{Fermi National Accelerator Laboratory, P.O. Box 500, Batavia, IL 60510, USA}
\affil{Kavli Institute for Cosmological Physics, University of Chicago, Chicago, IL 60637, USA}
\affil{Department of Astronomy and Astrophysics, University of Chicago, Chicago IL 60637, USA}

\author[0000-0001-8245-779X]{Catherine Fielder}
\UA%\affiliation{Steward Observatory, University of Arizona, 933 North Cherry Avenue, Rm. N204, Tucson, AZ 85721-0065, USA}

\author[0000-0003-3835-2231]{David~Mart\'{i}nez-Delgado}
\affiliation{Centro de Estudios de F\'isica del Cosmos de Arag\'on 
(CEFCA), Unidad Asociada al CSIC, Plaza San Juan 1, 44001 Teruel, Spain}

\affiliation{ARAID Foundation, Avda. de Ranillas, 1-D, E-50018 
Zaragoza, Spain}

\author[0000-0002-9144-7726]{Clara~E.~Mart\'inez-V\'azquez}
\affiliation{International Gemini Observatory/NSF NOIRLab, 670 N. A'ohoku Place, Hilo, Hawai'i, 96720, USA}

\author[0000-0002-0956-7949]{Kristine Spekkens}
%\affiliation{Department of Physics and Space Science, Royal Military College of Canada,\\ P.O. Box 17000, Station Forces Kingston, ON K7K 7B4, Canada}
\affiliation{Department of Physics, Engineering Physics and Astronomy, Queen’s University, Kingston, ON K7L 3N6, Canada}

\author[0000-0001-9775-9029]{Amandine~Doliva-Dolinsky}
\affil{Department of Physics \& Astronomy, University of Tampa, 401 West Kennedy Boulevard, Tampa, FL 33606, USA}
\affiliation{Department of Physics and Astronomy, Dartmouth College, 6127 Wilder Laboratory, Hanover, NH 03755, USA}

\author[0000-0001-5368-3632]{Laura C. Hunter}
\affiliation{Department of Physics and Astronomy, Dartmouth College, 6127 Wilder Laboratory, Hanover, NH 03755, USA}

\author[0000-0002-3936-9628]{Jeffrey L. Carlin}
\affiliation{AURA/Rubin Observatory, 950 North Cherry Avenue, Tucson, AZ 85719, USA}

\author[0000-0003-1697-7062]{William~Cerny}
\affiliation{Department of Astronomy, Yale University, New Haven, CT 06520, USA}

\author[0009-0005-9382-1362]{Tehreem N. Hai}
\affiliation{Department of Physics and Astronomy, Rutgers, the State University of New Jersey, 136 Frelinghuysen Road, Piscataway, NJ 08854, USA}

\author[0000-0001-5538-2614]{Kristen B.W. McQuinn}
\affiliation{Space Telescope Science Institute, 3700 San Martin Drive, Baltimore, MD 21218, USA}
\affiliation{Department of Physics and Astronomy, Rutgers, the State University of New Jersey, 136 Frelinghuysen Road, Piscataway, NJ 08854, USA}

 \author[0000-0002-6021-8760]{Andrew~B.~Pace}
\affiliation{Department of Astronomy, University of Virginia, 530 McCormick Road, Charlottesville, VA 22904 USA
}

\author[0000-0003-2599-7524]{Adam Smercina}\thanks{Hubble Fellow}
\affiliation{Space Telescope Science Institute, 3700 San Martin Drive, Baltimore, MD 21218, USA}

\begin{abstract}
We report the discovery of three faint and ultra-faint dwarf galaxies  -- Sculptor A, Sculptor B and Sculptor C -- in the direction of NGC~300 (D=2.0 Mpc), a Large Magellanic Cloud-mass galaxy. Deep ground-based imaging with Gemini/GMOS resolves all three dwarf galaxies into stars, each displaying a red giant branch indicative of an old, metal-poor stellar population. No young stars or \hi \ gas are apparent, and the lack of a {\it GALEX} UV detection suggests that all three systems are quenched. Sculptor C (D=2.04$^{+0.10}_{-0.13}$ Mpc; $M_V$=$-$9.1$\pm$0.1 mag or $L_V$=(3.7$^{+0.4}_{-0.3}$)$\times$10$^5$ $L_{\odot}$) is consistent with being a satellite of NGC~300. Sculptor A (D=1.35$^{+0.22}_{-0.08}$ Mpc; $M_V$=$-$6.9$\pm$0.3 mag or $L_V$=(5$^{+1}_{-1}$)$\times$10$^4$ $L_{\odot}$) is likely in the foreground of NGC~300 and at the extreme edge of the Local Group, analogous to the recently discovered ultra-faint Tucana B in terms of its physical properties and environment. Sculptor B (D=2.48$^{+0.21}_{-0.24}$ Mpc; $M_V$=$-$8.1$\pm$0.3 mag or $L_V$=(1.5$^{+0.5}_{-0.4}$)$\times$10$^5$ $L_{\odot}$) is likely in the background, but future distance measurements are necessary to solidify this statement. It is also of interest due to its quiescent state and low stellar mass. Both Sculptor A and B are $\gtrsim$2-4 $r_{vir}$ from NGC~300 itself.  The discovery of three dwarf galaxies in isolated or low-density environments offers an opportunity to study the varying effects of ram pressure stripping, reionization and internal feedback in influencing the star formation history of the faintest stellar systems.
\end{abstract}


\section{Introduction} \label{sec:intro}

The faintest galaxies are essential proving grounds for understanding dark matter and astrophysics on small scales \citep[][for recent reviews]{Bullock17,Simon19,Sales22}.  A hallmark of the $\Lambda$ Cold Dark Matter ($\Lambda$CDM) model is that structure forms hierarchically: galaxies inhabit dark matter halos, which contain dark matter substructures that often host smaller galaxies. Understanding this hierarchical structure formation in detail at sub-galactic scales has led to continuing, intensive efforts to observe and quantify the satellite system of the Milky Way \citep[e.g.][]{Drlica20,Cerny23,Smith23,Smith24,Gatto24,Homma24}, M31 \citep[e.g.][]{Martin13,DD22,DD23} and other nearby Milky Way-like galaxies \citep[e.g.][]{Chiboucas13,Crnojevic16,Crnojevic19,Smercina18,Bennet19,Bennet20,Carlsten22,Mao21,Mao24,Mutlu24}.  Meanwhile, numerical simulations that include baryonic astrophysics have made significant progress in reproducing the  dwarf galaxy properties of bright systems observed in Milky Way-mass systems  \citep[e.g.][]{Brooks13,Sawala16,Wetzel16,Samuel20,Engler21}, while semi-analytic models are successful for fainter systems \citep[e.g.][]{Manwadkar22,Weerasooriya23,Ahvazi24}.

To complement and extend observations around Milky Way-mass systems, and to test the astrophysics input into modern simulations, it is necessary to identify and understand the lowest-mass dwarf galaxies in a variety of environments. 
One area of interest is the population of dwarfs in lower density environments: at the edge of the Local Group's main galaxies \citep[e.g.][]{McQuinn23,McQuinn24}, around less massive hosts such as Magellanic Cloud-like systems \citep[e.g.][]{Sand15,Carlin16,Carlin21,Carlin24,McNanna24,Davis24}, and in isolation \citep[e.g.][]{Cannon11,McQuinn14,McQuinn21,Sand22,Jones23,Jones24,Li24}.  The effects of tidal and ram pressure stripping will be diminished in low density environments \citep[e.g.][]{Garling24}, with the exception of `backsplash' systems (e.g. \citealt{Teyssier12,Buck19,Santos23}), and so offer an opportunity to cleanly study other quenching mechanisms, such as reionization \citep{Bullock00,Benson02,Ricotti05,Jeon17,Applebaum21} and star formation/supernova feedback \citep[e.g.][]{Dekel86,Maclow99,Hopkins12,Elbadry18}.

Here we report the discovery of three faint, seemingly quenched, dwarf galaxies in the direction of NGC~300.  NGC~300 is an SA spiral galaxy at D=2.0 Mpc \citep[based on the tip of the red giant branch;][]{angst} in the direction of the Sculptor group.  It has a K-band luminosity very similar to the Large Magellanic Cloud \citep[][]{Mutlu_sims}, making it an excellent comparison for satellite population studies\footnote{{If we use a $K$-band mass-to-light ratio of 1, both NGC~300 and the LMC have a stellar mass of $\sim$2.6$\times$10$^9$ $M_{\odot}$. This is consistent with stellar mass estimates of 2$\times$10$^{9}$ $M_{\odot}$ for NGC~300 \citep{Munoz15} and 2.7$\times$10$^9$ $M_{\odot}$  for the LMC \citep{vdMarel02} in the literature.}}.  Throughout this work, we assume that NGC~300 has a virial radius of $r_{vir}$$\approx$120 kpc and halo mass of log($M_{halo}$/$M_{\odot}$)$\approx$11.29 as calculated in \citet{Mutlu_sims}, using the stellar mass--halo mass relation of \citet{Moster10}. These values are uncertain, but are used to help ascertain the relationship between the three faint dwarf discoveries and NGC~300, with this caveat in mind. We present the discovery of the three dwarf galaxies in Section~\ref{sec:discovery}, and discuss follow-up optical observations in Section~\ref{sec:data}. In Section~\ref{sec:props} we measure the basic physical properties of the dwarfs (distance, stellar population, gas content, structure and luminosity).  We discuss the environment of each in Section~\ref{sec:discussion} and compare them with other dwarfs in low density environments.
Finally, we summarize and conclude in Section~\ref{sec:summary}.

\begin{figure*}
\centering
\includegraphics[width=17.8cm]{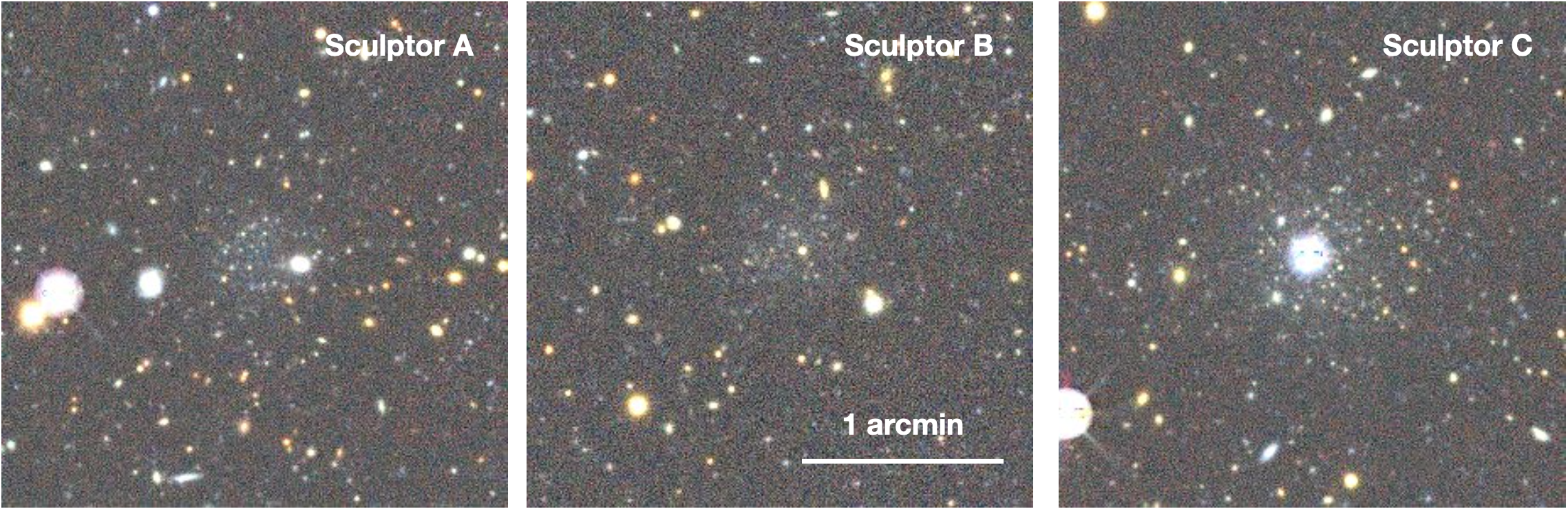}
\caption{Sculptor A, B and C as seen in the DESI Legacy Imaging Surveys sky browser.  North is up and east is to the left.  Note the bright foreground star projected onto Sculptor C. \label{fig:image}}
\end{figure*}

\section{Ultra-faint Dwarf Galaxy Discovery Around NGC~300} \label{sec:discovery}

Following the recent discovery of Tucana~B through a focused visual search utilizing the DESI Legacy Imaging Surveys and its interactive color image viewer\footnote{\url{https://www.legacysurvey.org/viewer}} \citep{Sand22}, we began a systematic search for faint companions around other nearby galaxies (D$\lesssim$2 Mpc), including NGC~300 (for reference, Data Release 9 of the DESI Legacy Imaging Surveys was used).  We uploaded a custom file to mark off a region with a projected box size of 400 kpc on a side centered on NGC~300 itself ($\approx$11.3 deg), and visually searched for stellar over-densities with underlying diffuse light. This coverage area out to $R_{proj}$$\sim$200 kpc encloses the full virial radius of NGC~300, which is approximately 120 kpc \citep{Mutlu_sims}. Faint dwarf galaxies at the distance of NGC~300 can resemble `semi-resolved' objects, with both resolved and diffuse components, at the depth of DECaLS ($g,r$$\approx$23.5-24 mag; depending on the field; \citealt{Dey19}). We searched the NGC~300 footprint several times, at different spatial scales and contrast levels.

\begin{figure}
%\centering
\includegraphics[width = 0.95\linewidth]{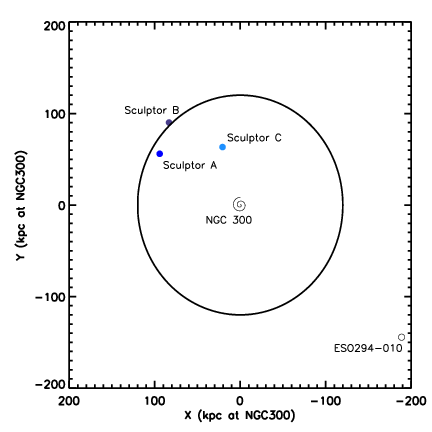}
\caption{Spatial footprint of our visual search for dwarfs associated with NGC~300.  The { solid} line marks the approximate projected virial radius at $R_{vir, NGC~300}$=120 kpc \citep{Mutlu_sims}, and the unfilled point shows the position of ESO294-010 (D=1.94 Mpc; M$_B$=$-$10.9 mag; \citealt{angst}), a previously known dwarf galaxy which may be associated with NGC~300.  To our knowledge, these are the only known galaxies within the virial region of NGC~300, although the galaxy NGC~55 (also at $D$=2 Mpc) is directly to the west of this field. The spatial position of Sculptor A, B and C are shown. 
Sculptor A (D=1.35$^{+0.22}_{-0.08}$ Mpc) and Sculptor B (D=2.48$^{+0.21}_{-0.24}$ Mpc) are both at the approximate virial radius of NGC~300 in projection, and are $\sim$500 kpc offset in distance as well.  We discuss the membership status of the identified dwarfs further in Section~\ref{sec:env}. 
\label{fig:footprint}}
\end{figure}

Three high-confidence faint dwarf galaxy candidates stood out during the search, and we show color cutouts obtained from the Legacy Survey Image Viewer in Figure~\ref{fig:image}. We also present a schematic of the footprint searched, and the projected position of these discoveries in Figure~\ref{fig:footprint}. NGC~300 and surrounding areas are in the footprint of the Dark Energy Survey Data Release 2 \citep[DES DR2; ][]{DES_DR2}, and we downloaded photometry of the field using NOIRLab's Query Interface Tool\footnote{\url{https://datalab.noirlab.edu/query.php}}.  This photometry was suggestive of an old, metal-poor red giant branch (RGB) at $\sim$2 Mpc but was ultimately inconclusive.  We thus sought deeper ground-based optical imaging, which we discuss next.

We choose the names Sculptor A, Sculptor B and Sculptor C for the three new faint dwarfs because of the constellation they reside in, and the fact that we are uncertain if each of these dwarfs is associated with NGC~300 (as we address in later sections of this paper).  A similar naming convention has been used for other dwarfs just beyond the edge of the Local Group.

\section{Gemini Deep Optical Follow-Up}\label{sec:data}

%Gemini
%2x2 binning

After the visual discovery of Sculptor A, B and C, we obtained deep follow-up imaging of all three systems with the Gemini South telescope using the Gemini Multi-Object Spectrograph  \citep[GMOS;][]{GMOS} under the Fast Turnaround program GS-2022B-FT-103.  The GMOS images have a $\approx$5.5\arcmin$\times$5.5\arcmin \ field of view and 0.16 arcsec/pixel scale after binning.  Both $g$ and $r$-band imaging was taken for all three dwarfs, with strict image quality constraints, between 2022 August 29 and 2022 September 30 (UT).  For Sculptor A and B, we took 7$\times$300s exposures in both bands.   For Sculptor C, we obtained 16$\times$120s exposures in both bands in order to mitigate the effects of the bright foreground star centered on the dwarf (see Figure~\ref{fig:image}).  Small dithers were taken between exposures.  At the time of the observations, GMOS South suffered from a bad amplifier, which we were careful to avoid in our observational setup.  All data associated with this amplifier was masked during the data reduction process.

Initial data reduction for the Gemini data was done with DRAGONS \citep{dragonsRNAAS_2023}, the pipeline maintained by Gemini Observatory.  DRAGONS performs bias subtraction, flat-field correction and bad pixel masking on the images. Cosmic rays were rejected using the sigma-clipping method within the DRAGONS pipeline. Stacked images were created using the weighted average of the constituent images.  As a final step, an astrometric correction was applied by using {\sc scamp} \citep{scamp}.  The final $g$ and $r$ band stacked images had point spread function full width half maximum values between 0.6\arcsec \ and 0.8\arcsec.

We performed point-spread function fitting photometry on the stacked GMOS images, using {\sc daophot} and {\sc allframe} \citep{Stetson87,Stetson94}, following the general procedure described in \citet{Mutlu18}.  The photometry was calibrated to point sources in the DES DR2 catalog \citep{DES_DR2}, including a color term, and was corrected for Galactic extinction \citep{Schlafly11} on a star by star basis.  The typical color excess at the position of the dwarfs is  $E(B-V)$=0.014--0.015 mag (see Table~\ref{tab:props}).  In the remainder of this work, we present dereddened $g_0$ and $r_0$ magnitudes, unless otherwise stated.

To determine our photometric errors and completeness as a function of magnitude and color, we conduct artificial star tests with the {\sc daophot} routine {\sc addstar}, similar to previous work \citep{Mutlu18}.  Over several iterations, we injected $\sim$10$^5$ artificial stars into our stacked images with a range of magnitudes ($r$=18--29 mag) and colors ($g-r$=$-$0.5 to 1.5), and then photometered the simulated data in the same way as the original images. The 50\% (90\%) completeness level was at $r$= 26.4, 26.3, 26.4 (25.4, 25.1, 25.5) and $g$=26.7, 26.7, 27.0 (25.5, 25.5, 25.9) mag for Sculptor A, B and C respectively.  

\begin{figure*}
\centering
\includegraphics[width=10.cm]{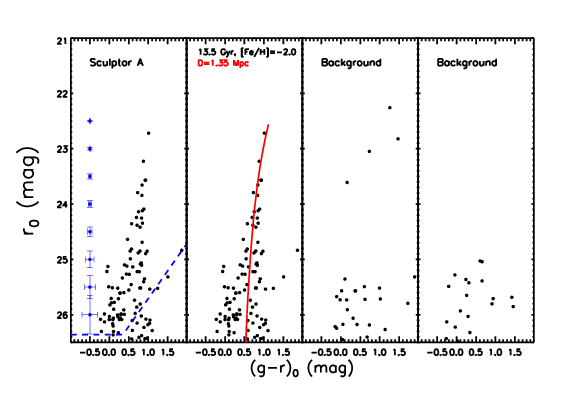}\vspace{-0.7cm}
\vspace{-0.7cm}
\includegraphics[width=10.cm]{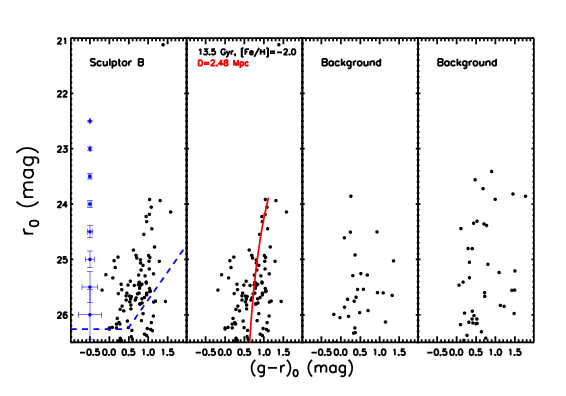}
\includegraphics[width=10.cm]{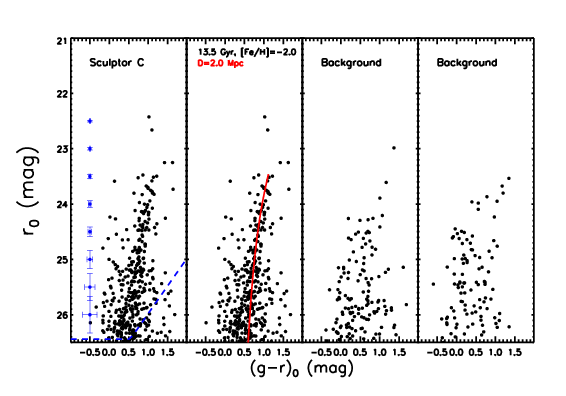}\vspace{-0.35cm}
\caption{Deep CMDs for our three dwarf galaxies based on Gemini GMOS imaging data. For each plot, the left two panels display the CMD of the dwarf within 1.0 $r_h$ (Section~\ref{sec:struct}).  Only point sources are plotted, as described in Section~\ref{sec:data}. The far left panel shows the typical uncertainties at different $r$-band magnitudes, while the dashed line marks the 50\% completeness limit, both as determined by artificial star tests.  The center-left panel overplots a 13.5 Gyr, [Fe/H]=$-$2.0 isochrone \citep{Dotter08} at the distance determined by our CMD-fitting technique (Section~\ref{sec:dist}).  The two panels on the right show randomly selected background regions, chosen to be devoid of bright, saturated stars.  The background regions have different areal coverage for each dwarf so that they are equivalent to that covered by the half light radius.    \label{fig:CMD}}
\end{figure*}

In Figure~\ref{fig:CMD} we show the color-magnitude diagrams (CMDs) of our three dwarfs within their half-light radius, $r_h$,  as derived in Section~\ref{sec:struct}. As can be seen, each dwarf has a clear but sparsely populated red giant branch whose brightest stars are at different magnitudes from dwarf to dwarf, likely indicating that these systems are at different distances. %, along with several equal area “background” CMDs. 
We discuss the structure and stellar populations of these new dwarfs in the following section.

\begin{deluxetable*}{lccc}
\centering
    \tablecaption{Faint Dwarf Galaxy Properties \label{tab:props}}
\tablehead{\colhead{Parameter} & \colhead{Sculptor~A} & \colhead{Sculptor~B} & \colhead{Sculptor~C}}
\startdata
$\alpha_0$ (J2000) & 01:08:30.8 $\pm$ 2.4" & 01:06:50.9 $\pm$ 3.6" & 00:57:52.2 $\pm$ 3.1"\\
$\delta_0$ (J2000) & $-$36:03:52.9 $\pm$ 3.2" & $-$35:04:39.0 $\pm$ 2.7" & $-$35:51:08.4 $\pm$ 4.0" \\
$E(B-V)$ (mag) & 0.015 & 0.015 & 0.014\\
$m-M$ (mag) & 25.65$^{+0.33}_{-0.13}$ & 26.98$^{+0.17}_{-0.23}$ & 26.55$^{+0.10}_{-0.13}$\\
Distance (Mpc) & 1.35$^{+0.22}_{-0.08}$ & 2.48$^{+0.21}_{-0.24}$ & 2.04$^{+0.10}_{-0.11}$ \\
R$_{\rm proj, NGC300}$ (kpc)$^a$ & 109 & 123 & 67\\
 D$_{\rm 3D, NGC300}$ (kpc)$^b$ & 660$^{+80}_{-220}$ &  500$^{+210}_{-230}$ & 80$^{+80}_{-13}$\\
 $M_V$ (mag) & $-$6.9$\pm$0.3 & $-$8.1$\pm$0.3 & $-$9.1$\pm$0.1 \\
 $L_V$ ($L_\odot$) & (5$^{+1}_{-1}$)$\times$10$^4$ & (1.5$^{+0.5}_{-0.4}$)$\times$10$^5$ & (3.7$^{+0.4}_{-0.3}$)$\times$10$^5$\\
 $\log (M_{*}/M_\odot)$ & 4.7$\pm$0.1 & 5.1$\pm$0.1 & 5.7$\pm$0.1 \\
 $r_h$ (arcsec) & 16.8 $\pm$ 4.2 & 20.4 $\pm$ 7.1 & 36.6 $\pm$ 4.2 \\
 $r_h$ (pc) & 110$\pm$28 & 245$\pm$85 & 362$\pm$42 \\
 $\epsilon$ & 0.34 $\pm$ 0.17 & 0.41 $\pm$ 0.22 & 0.24 $\pm$ 0.10 \\
 $\theta$ (deg) & $-$11.5 $\pm$ 17.5 & 133.1 $\pm$ 29.2 & 66.9 $\pm$ 12.6 \\
 $\log (\mathrm{SFR_{NUV}/M_\odot \, yr^{-1}})$ & $<$6.0 & $<$5.3 & $<$4.8\\
 $\log (\mathrm{SFR_{FUV}/M_\odot \, yr^{-1}})$ & $<$6.3 & $<$5.6 & $<$6.0\\
 $\log (M_\mathrm{HI}/M_\odot)$ & $<$5.3 & $<$5.8 & $<$5.6\\
\enddata
\tablenotetext{a}{Projected radius with respect to NGC~300 (D=2 Mpc).
\tablenotetext{b}{3D distance between dwarf galaxy and NGC~300 (D=2 Mpc).}}
\end{deluxetable*}

%\footnote{\url{https://www.legacysurvey.org/viewer}}

\section{Dwarf Galaxy Physical Properties} \label{sec:props}

In this section we measure the physical properties of the three newly discovered dwarf galaxies, using the deep Gemini photometry, as well as archival \hi \ and {\it GALEX} UV data sets.  The properties we derive are shown in Table~\ref{tab:props}.

\subsection{Distance}\label{sec:dist}

As we will discuss in the next section, the stellar populations of all three dwarf galaxies exhibit sparsely populated old, metal-poor red giant branches (RGBs), with no other stellar population apparent (see Figure~\ref{fig:CMD}).  Because of this sparse population, with few stars populating the brightest regions of the RGB, only one of the three dwarfs displays a clear tip of the red giant branch (TRGB) which could be used for measuring a distance (Sculptor~C, see below).  The Gemini data is also not deep enough to obtain a distance based on the horizontal branch or RR Lyrae stars; although the first signs of a horizontal branch may be visible in Sculptor~A at the faintest magnitudes, as we discuss below.

For consistency, we thus measure the distance to the three dwarf galaxies using a CMD-fitting technique, comparing the number of stars consistent with an old, metal-poor theoretical isochrone as a function of assumed distance.  A similar technique has been used to measure the distance to many Milky Way ultra-faint dwarf galaxies \citep[e.g.][]{Walsh08,Sand09}, as well as Tucana B \citep{Sand22}.  In our analysis we use Dartmouth isochrones \citep{Dotter08} with a 13.5 Gyr stellar population and low metallicities ([Fe/H]=$-$2.5, $-$2.0 and $-$1.5.).  For each dwarf we include all stars brighter than the 50\% completeness threshold, and stars spatially coincident with the main body of each dwarf (a radius of 0.3--0.5\arcmin), assessed visually.  Then, we shift each isochrone fiducial through distance moduli ($m-M$) between 25.0 and 28.0 mag (1.0 and 4.0 Mpc) in 0.025 mag steps, counting the number of stars consistent with the isochrone at that distance.  A scaled background region at the edge of the GMOS field of view is used to subtract off contaminants for each distance modulus trial.  For a given $r_0$, a star is considered consistent with the isochrone if its $(g-r)_0$ color is within a red/blue boundary derived from our photometric uncertainties.  The best distance corresponds to the distance modulus which maximizes the number of stars consistent with the isochrone after background removal.  { To assess our uncertainties on the distance measurement, we bootstrap (resample with replacement) both the input dwarf stars, and the background objects.}

  The fits to the [Fe/H]=$-$2.0 isochrone are excellent, as are those for the [Fe/H]=$-$2.5 isochrone, while the best-fitting [Fe/H]=$-$1.5 isochrone are slightly too red and do not capture the shape of the observed RGB.  For this reason, we adopt the [Fe/H]=$-$2.0 distance in this work (see the second-left panel in Figure~\ref{fig:CMD}), with the range of best-fitting distance values for the full metallicity set as our formal uncertainty.  { We choose this metric for the reported uncertainty rather than the results of the bootstrap analysis, as the bootstrap uncertainties are sub-dominant in comparison to the distance spread between metallicities.}  We list these distances in Table~\ref{tab:props}.  

As a check on our CMD-fitting results, we also found the distance to Sculptor~C using a standard TRGB distance measurement technique. Here we selected stars in the dwarf consistent with the red giant branch and measured the luminosity function of these stars, compared with a model luminosity function convolved with our photometric uncertainties and completeness as determined from our artificial star tests \citep[see][for further details]{Crnojevic19}.  Using a nonlinear least squares fit between model and data, we calculated a TRGB magnitude of $r_{TRGB}$=23.58$\pm$0.08 mag.  This results in a distance modulus of $m-M$=26.59$\pm$0.13 mag (assuming $M_{r}^\mathrm{TRGB} = -3.01 \pm 0.01$, which does not include systematic or zeropoint uncertainties; \citealt{Sand14}), and a distance of $D = 2.08\pm0.12$ Mpc.  This value is nearly identical to that determined through the CMD-fitting technique described above, so we stick with the original measurement for all three dwarfs, for consistency.

Interestingly, the distances to Sculptor A (D=1.35$^{+0.22}_{-0.08}$ Mpc) and Sculptor B (D=2.48$^{+0.21}_{-0.24}$ Mpc) are not clearly consistent with the distance to NGC~300 itself (D=2.0 Mpc), especially given that Sculptor A and Sculptor B are roughly at the projected virial radius of $\approx$120 kpc.  We return to this and the membership status of the dwarfs in Section~\ref{sec:env}.

\subsection{Stellar Population}\label{sec:pop}

The CMDs of all three dwarf galaxies appear to only consist of an old, metal-poor stellar population (Figure~\ref{fig:CMD}).  Any younger, blue stellar population either has few stars associated with it, or is below our detection limit. Many of the recently discovered faint dwarf galaxies beyond the Local Group show distinct signs of recent star formation (e.g., Leo~P, \citealt{McQuinn15}; Antlia~B, \citealt{Hargis20}; Pavo, \citealt{Jones23}), although a growing subset also appears to be quenched, with little to no recent star formation (e.g. Tucana~B, \citealt{Sand22}; Blobby, \citealt{Casey23}; Hedgehog, \citealt{Li24}).  The mix of stellar populations of faint dwarf galaxies in the `field' is a critical ingredient for understanding the role of reionization, stellar feedback and ram pressure from the cosmic web in driving the evolution of the smallest galaxies.

The distance to Sculptor A is near enough (D=1.35 Mpc) that we could plausibly have detected blue horizontal branch stars.  Given the distance, such a population would reside at $r_0$$\approx$26.3 mag and $(g-r)_0$$\lesssim$0 mag.  Visual inspection of the CMD in Figure~\ref{fig:CMD} indicates that there may be an overdensity of stars in this region of color-magnitude space, although the contamination is high and photometric uncertainties are significant. Space-based observations are necessary to confirm this population.  

To constrain any possible young stellar population in each dwarf, we search for coincident UV emission with the {\it Galaxy Evolution Explorer} ({\it GALEX}; \citealt{galex})  data archive.  {\it GALEX} data from the All Sky Survey are available for Sculptor A and Sculptor B, while data from the Medium Imaging Survey are available for Sculptor C (see \cite{Bianchi17} for details). {\it GALEX} observations are sensitive to star formation on $\lesssim$100 Myr timescales \citep{Lee11}.  For our measurements, we adopt the methodology of \citet{Karunakaran21}, using an aperture 1.33$r_h$ in size (see next section), and placing down 1000 random apertures throughout the {\it GALEX} field to assess flux uncertainties (after masking bright objects).  No flux was detected for any of the three dwarf galaxies, and we use the {\it GALEX} NUV and FUV flux limits to determine star formation rate limits using the relations of \citet{Iglesias-Paramo+2006}. See Table~\ref{tab:props} for our 2-$\sigma$ star formation rate limits for each object.  The star formation limits are $\log (\mathrm{SFR_{NUV}/M_\odot \, yr^{-1}}) < -4.8$ and $\log (\mathrm{SFR_{FUV}/M_\odot \, yr^{-1}}) < -5.6$ for all three dwarfs, which are more stringent than most UV detections in satellite galaxies around Milky Way-like galaxies \citep[e.g.][]{Karunakaran21}.  While there is no recent star formation among the dwarfs, future space-based observations are necessary to characterize any faint, intermediate-age population ($\gtrsim$500 Myr; e.g. \citealt{Weisz14M31}).

\subsection{Stellar Structure, Luminosity \& Stellar Mass}\label{sec:struct}

To measure the structural parameters of these sparse stellar systems, we fit an exponential profile to the two-dimensional distribution of stars consistent with the RGB in each system, using the maximum likelihood technique of \citet{Martin08}, as implemented in \citet{Sand12}.  Stars throughout the GMOS field that are consistent with the best-fitting Dartmouth isochrone and distance are used in the analysis (see Section~\ref{sec:dist}).  The free parameters for the exponential fit are the central position ($\alpha_0$, $\delta_0$), position angle ($\theta$), ellipticity $\epsilon$, half-light radius ($r_h$) and a constant background surface density.  Uncertainties are calculated using a bootstrap resampling analysis, with 1000 iterations.  The maximum likelihood technique employed naturally accounts for portions of the image where point sources would be undetectable, such as the bad amplifier (mentioned in Section~\ref{sec:data}) and the bright star coincident with the main body of Sculptor C (see Figure~\ref{fig:image}).  
The results of the structural analysis are presented in Table~\ref{tab:props}.  The dwarfs are $\approx$100-360 pc in size (as characterized by the half-light radius) and have moderate ellipticities ($\epsilon$$\approx$0.2--0.4) which are all typical values for faint dwarf galaxies, as we will discuss in Section~\ref{sec:discussion}.

To measure the luminosity of the three dwarf galaxies, we employ the technique of \citet{Martin08} in the ``CMD shot noise'' regime, when the presence or absence of individual stars in the upper regions of the RGB can significantly affect the overall luminosity.  To do this, we produce a well-populated CMD using a 13.5 Gyr, [Fe/H]=$-$2 stellar population with a Salpeter initial mass function, convolving the stellar population with the photometric uncertainties and completeness derived from our artificial star tests.  From this simulated CMD we randomly draw the same number of stars as was found in our maximum likelihood analysis (which can account for missing areas of data such as the bright star in Sculptor C), and to this luminosity we add the luminosity in the simulated population below our detection limit.  We repeat this process 100 times, using the median and standard deviation as our final absolute magnitude and uncertainty, including the distance uncertainty, into the final estimation.  We convert to $V$-band magnitudes using the filter transformation of \citet{Jordi06}.  Absolute magnitudes and luminosities in the $V$-band are quoted in Table~\ref{tab:props}.

Finally, to estimate the stellar mass of each object, we use the $g-r$ color and magnitudes to determine the stellar mass-to-light ratio using the \citet{Into13} relation.  The stellar masses closely track the V-band luminosity, as expected for old metal poor stellar populations; see Table~\ref{tab:props} for our derived values.

\begin{figure*}
\centering
\includegraphics[width = 0.45\linewidth]{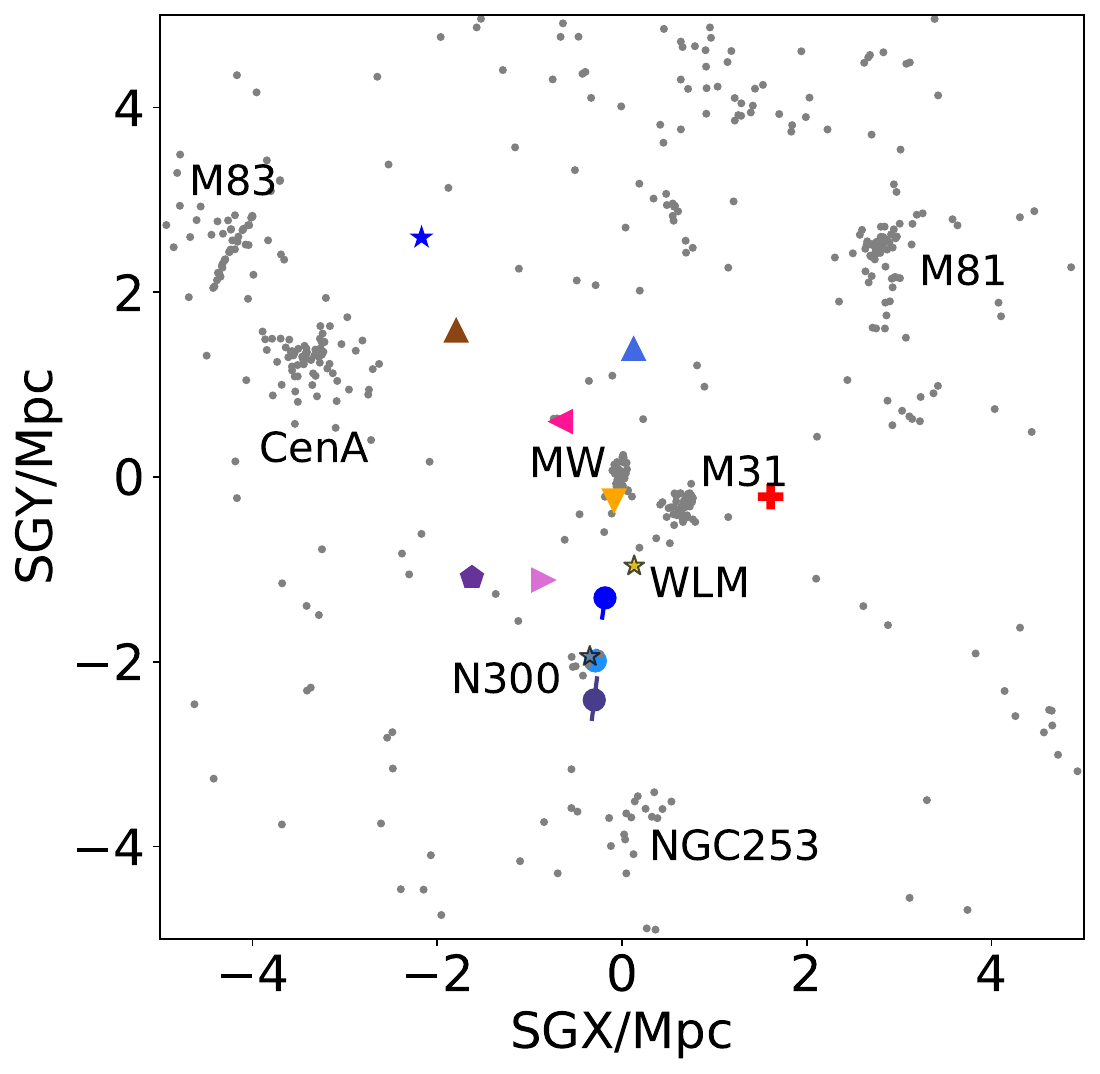}
\includegraphics[width = 0.45\linewidth]{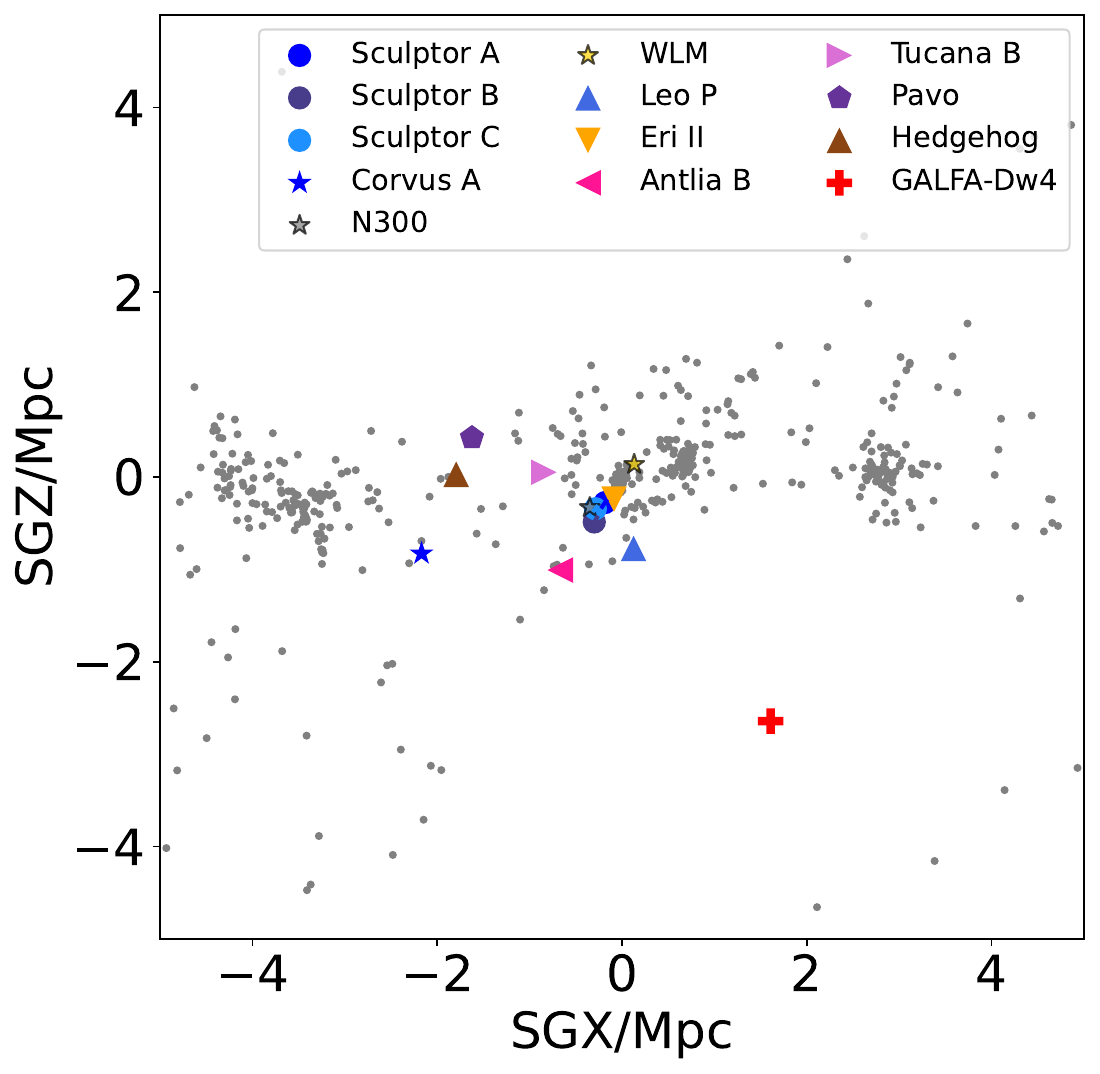}
\caption{A map of nearby galaxies in supergalactic coordinates, illustrating the environment of Sculptor A, B and C.  Left: The supergalactic $X-Y$. Sculptor A, B and C are shown, along with their appropriately scaled distance uncertainties.  It is clear that Sculptor A and Sculptor B are in front of and behind NGC~300, respectively, in this projection. We show several other dwarfs in low density or isolated environments for reference: Corvus~A \citep{Jones24}; Leo~P \citep{Giovanelli13}; Eridanus II \citep{Crnojevic16}; Antlia~B \citep{Sand15}; GALFA-Dw4 \citep{Bennet22}; Tucana B \citep{Sand22}; Pavo \citep{Jones23}; and  Hedgehog \citep{Li24}.  Grey points are taken from the compilations of \cite{Kara04,Kara19}. We also mark NGC~300 and the WLM galaxy, which are both mentioned in the text. Right: We show the supergalactic $X-Z$ plane and the local plane of galaxies.
\label{fig:environment}}
\end{figure*}

\subsection{Gas Content}\label{sec:gas}

None of the three dwarfs are detected in \hi \ line emission in the \hi \ Parkes All Sky Survey \citep[HIPASS;][]{Barnes+2001}, suggesting that they are not gas-rich. To determine upper limits on the \hi \ masses of the dwarfs we use the Galactic All Sky Survey \citep[GASS;][]{McClure-Griffiths09,Kalberla10,Kalberla15}, which is more sensitive than HIPASS, but only has data within the velocity range $-500 < v/\mathrm{km\,s^{-1}} < 500$. Given the radial velocity of NGC~300 (140 $\mathrm{km\,s^{-1}}$) and the distance estimates to the three dwarfs, this range is likely more than sufficient, but a caveat to our upper limits below is that they would not apply for $v > 500$~\kms. We also cannot be certain (without a prior velocity measurement) that \hi \ line emission from the dwarfs is not hidden within MW emission.

We downloaded GASS spectral line cubes\footnote{\url{https://www.astro.uni-bonn.de/hisurvey/gass/index.php}} for each dwarf and extracted a spectrum over $3\times3$ spatial pixels, centered on each target. We inspected these spectra (and the cubes) at spectral resolutions of 5 and 25~\kms, but saw no significant emission features other than the MW and high-velocity clouds. We measured the rms noise at 25~\kms \ resolution for each dwarf and assuming 25~\kms \ as a conservative fiducial velocity width, estimated the $3\sigma$ upper limits in each case. These limits are given in Table~\ref{tab:props}.  More stringent gas limits will be necessary to verify that these systems are truly gas free \citep[down to $M_{HI}$/L$_V$$\approx$1; e.g.][]{Putnam21}.

\section{Discussion} \label{sec:discussion}

The three dwarf galaxies in this work are amongst the faintest quenched dwarfs discovered outside the Local Group.  In this section we discuss whether they are plausible satellites of NGC~300, and their environment more generally.  From there, we compare the dwarfs to other systems in low-density environments outside the Local Group and Magellanic Cloud-analogs.

\subsection{Environment \& NGC~300 Membership Status} \label{sec:env}

The original goal of our visual search was to identify satellites of the LMC-analog NGC~300.  However, based on the measured distances and projected radii of our three new dwarfs, we must first assess whether they are likely satellites or foreground/background systems.

We plot the projected radial distribution of the three dwarfs with respect to NGC~300 in Figure~\ref{fig:footprint}, along with the only other known galaxy within the extended virial region of NGC~300, ESO294-010. First, two out of three of the dwarfs (Sculptor A and Sculptor B) are near the projected virial radius of NGC~300, with $R_{proj, NGC~300}$=109 and 123 kpc, respectively.  The third dwarf, Sculptor C, has a projected radius of $R_{proj, NGC~300}$=67 kpc, and given it has a distance consistent with NGC~300 itself, we confidently assert that it is very likely to be a satellite.  

Not only are Sculptor A and B in the projected outskirts of NGC~300, their distances also indicate that they may not be associated with the galaxy.  At $\approx$1.35 Mpc, Sculptor A has a 3D distance of D$_{\rm 3D, NGC300}$=660$^{+80}_{-220}$ kpc, meaning it is at least $\gtrsim$3.7 $r_{vir, NGC~300}$ assuming a nominal $r_{vir, NGC~300}$=120 kpc.   Similarly, Sculptor B (at D$\approx$2.48 Mpc) has a 3D distance from NGC~300 of D$_{\rm 3D, NGC300}$=500$^{+210}_{-230}$ kpc, or $\gtrsim$2.3 $r_{vir, NGC~300}$.  Based on these distances, we suggest that both systems are not associated with NGC~300.

Despite the assessment that Sculptor~B is not associated with NGC~300, several caveats are worth mentioning.  First, it is still plausible that Sculptor B is a backsplash system of NGC~300, where the expectation is that such systems will be at $\lesssim$2.5 $r_{vir}$ \citep{Teyssier12,Diemer15,Buck19,Applebaum21}. Recent  work suggests that extreme backsplash systems at larger radii are possible; a future velocity measurement of Sculptor~B will help distinguish between scenarios \citep[see][for a discussion of the Tucana dwarf]{Santos23}.  It is also true that little theoretical/numerical work on the backsplash systems of Magellanic Cloud-scale halos has been done. Such studies will be useful as near-future discoveries of faint dwarfs become more common. Finally, it is also the case that the virial radius of NGC~300 is not directly measurable and highly uncertain, making an assessment of Sculptor~B's status even more difficult. %Sculptor~A is even farther from NGC~300.
%Further assessment of any possible association between NGC~300 and the three dwarfs should also consider the velocity of each with respect to NGC~300 and other local structures.  Those observations are a sensible next step.

To investigate the environment of the new Sculptor dwarfs further, we plot their positions (along with other Local Volume dwarfs) on two projections of the supergalactic coordinate system in Figure~\ref{fig:environment}.  On the plot, we also highlight other recently discovered nearby dwarf galaxies in low-density and isolated environments.  In this view, Sculptor A clearly appears in the foreground to NGC~300 and associated galaxies.  The WLM galaxy is marginally closer to Sculptor A then NGC~300 ($\approx$630 kpc), although they are consistent within the uncertainties.  In this sense, Sculptor A is very similar to Tucana~B, at a similar distance to the Milky Way and its nearest neighbor (IC~5152 at $\approx$620 kpc).  

The LMC-mass galaxy NGC~55 (D$\approx$2 Mpc) is a nearby neighbor to NGC~300, located to the west just outside the projected field of view of Figure~\ref{fig:footprint}. NGC~55 and NGC~300 are sometimes considered part of the same overall group of galaxies \citep[e.g.][]{Mcconachie12}.  Interestingly, all three newly discovered Sculptor dwarfs are arrayed on one side of NGC~300 (in projection), in the direction opposite to that of NGC~55.  Given the range of distances in the new Sculptor dwarfs, there is no evidence for planar structure, although future refined distance measurements are necessary.

The quenched Sculptor dwarfs join a rapidly growing list of nearby, relatively isolated dwarf galaxy systems which span a range of star formation properties. They offer an opportunity to understand what drives star formation and quenching in systems far from a massive perturbing galaxy. 

\subsection{Comparisons with Dwarfs in Low-Density Environments}

\begin{figure}
\centering
\includegraphics[width=8.75cm]{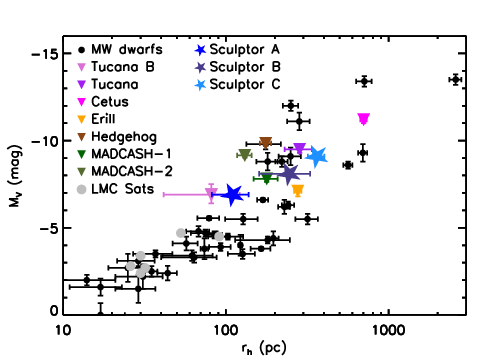}
\caption{Absolute magnitude as a function of half-light radius of Sculptor A, B and C.  We also plot a variety of quenched dwarf galaxies in low-density environments.  These include quenched dwarfs in the outskirts of the Local Group (Cetus: \citealt{McConnachie06}; Tucana: \citealt{Savian96}; Eri II: \citealt{Crnojevic_erii,Simon21}); in isolated environments (Tucana~B: \citealt{Sand22}; Hedgehog: \citealt{Li24}); around Magellanic Cloud analogs (MADCASH-1 \& 2; \citealt{Carlin21}); and the Large Magellanic Cloud itself (Reticulum II: \citealt{Mutlu18}; Phoenix II: \citealt{Mutlu18}; Horologium I: \citealt{Richstein24}; Carina II: \citealt{Torrealba18}; Carina III: \citealt{Torrealba18}, Hydrus I: \citealt{Koposov18}).   Also plotted are the dwarf satellites of the Milky Way for context (see the Local Volume database: \url{https://github.com/apace7/local_volume_database}).    \label{fig:mvrh}}
\end{figure}

We plot the size-luminosity relation of faint dwarf galaxies around the Milky Way in Figure~\ref{fig:mvrh}, along with several other quenched dwarfs in isolated or low-density environments.  These include the putative ultra-faint satellites of the LMC (Reticulum II, Phoenix II, Horologium I, Hydrus I, Carina II and Carina III; see e.g. \citealt{Battaglia22}), dwarfs associated with Magellanic Cloud analogs (MADCASH-1 \& MADCASH-2; \citealt{Carlin21}), several dwarfs at the edge of the Local Group (Tucana, Cetus, Eridanus II), and even more isolated systems (Tucana~B \& Hedgehog).
Overall, Sculptor A, B and C are consistent with the size-luminosity relation of faint dwarfs in these various environments, although surveys outside the Local Group do not have sensitivity to the ultra-faint dwarfs that are associated with the LMC.

Sculptor~C ($M_V$=$-$9.1 mag; $r_h$=362 pc) can be compared directly to recent faint dwarf galaxy discoveries around Magellanic Cloud analogs, given its very likely status as an NGC~300 satellite.  Here MADCASH-2 ($M_V$=$-$9.2; $r_h$=130 pc), associated with NGC~4214, is approximately the same luminosity but more compact.  MADCASH-1 ($M_V$=$-$7.8; $r_h$=180 pc; associated with NGC~2403) is a significantly fainter system \citep[physical parameters for both dwarfs are from][]{Carlin21}.  Interestingly, Hubble Space Telescope data of MADCASH-2 shows evidence for a small amount of recent star formation ($<$1.5 Gyr). Similar quality data will be necessary to further investigate small amounts of recent star formation in our dwarf sample.

The known satellites of the SMC-like galaxy NGC~3109 \citep[$M_V$$\approx$$-$15, D$\approx$1.28 Mpc;][]{Mcconachie12} offer another point of comparison.  Both Antlia ($M_V$=$-$10.4) and Antlia~B ($M_V$=$-$9.4 mag) have neutral gas reservoirs and notable recent star formation episodes \citep[e.g.][]{Weisz11,Sand15,Hargis20}, in contrast to Sculptor~C and the MADCASH dwarfs, although their greater masses may also contribute to their ability to retain gas in an SMC-like environment.  Going forward, building a sample of satellite properties in the SMC to LMC mass range may shed light on the effectiveness of ram pressure stripping around lower mass halos \citep[see][for a recent discussion]{Garling24}.

Sculptor~A is most similar to Tucana B in several respects, including the fact they they are the only two isolated, quenched ultra-faint dwarf galaxies known. Both systems are $\approx$1.4 Mpc from the Milky Way ($\gtrsim$4.5 $r_{vir,MW}$, assuming $r_{vir,MW}$=200-300 kpc), likely too distant to be backsplash systems \citep[e.g.][]{Teyssier12,Buck19,Applebaum21}.  They have very similar luminosities ($M_V$=$-$6.9; $L_V$=5$\times$10$^4$ $L_\odot$) and sizes within the uncertainties.  The two systems also share a lack of young stars and neutral hydrogen gas, although this needs to be confirmed by deeper follow-up observations.  Sculptor~A, like Tucana~B, likely has a low enough stellar mass to have been significantly affected by reionization and internal feedback in the early universe \citep[$M_*$$\lesssim$10$^{5-6}$ $M_{\odot}$, e.g.][]{Applebaum21}.  

Finally, Sculptor~B ($M_V$=$-$8.1 mag) appears to be a brighter, more distant version of Sculptor~A.  It is somewhat fainter than the Milky Way satellites Draco \citep[$M_V$=$-$8.7 mag;][]{Munoz18} and Canes Venatici I \citep[$M_V$=$-$8.8 mag;][]{Munoz18}, and significantly fainter than the Cetus and Tucana dwarfs, which are both suspected to be backsplash systems of the Milky Way.  If Sculptor~B is a distant backsplash system of NGC~300, it would be amongst the faintest examples of this class, and provide a unique opportunity to study such systems around an LMC-analog.  If, however, it is unassociated with NGC~300 it is another candidate to be strongly affected by reionization or stellar feedback.

\section{Conclusions \& Implications}\label{sec:summary}

We have presented the discovery of three faint and ultra-faint dwarf galaxies in the direction of NGC~300, an isolated, LMC-analog.  Deep Gemini GMOS follow-up imaging resolved each dwarf into stars, allowing us to determine their distance, luminosity, structure and basic star formation properties.  The dwarfs range in distance from $\approx$1.3-2.5 Mpc, and have absolute magnitudes ranging from the ultra-faint regime (Sculptor~A; $M_V$=$-$6.9 mag) to the faintest `classical' dwarfs ($M_V$=$-$8.1 and $-$9.1 mag for Sculptor B \& C, respectively). All three dwarfs appear quenched, with an exclusively old stellar population, and no detectable \hi \ gas reservoir or GALEX UV emission.  %Further deep HI observations and space-based imaging are necessary to further constrain these systems.

Interestingly, only one out of three of the dwarfs is clearly associated with NGC~300 itself, Sculptor~C (D=2.04 Mpc; $M_V$=$-$9.1 mag).  The presence of Sculptor~C is consistent with initial results from surveys searching for the faint satellite populations of Magellanic Cloud-analog systems \citep[e.g.][]{Carlin16,Carlin21,McNanna24}, although it is possible that the present search missed other dwarf galaxy candidates.  In the near future, the DECam Local Volume Exploration Survey (DELVE; \citealt{DELVEdr1,DELVEdr2}) will perform a systematic search for faint dwarf satellites around NGC~300 (and several other Magellanic Cloud-mass systems) in order to robustly constrain the population and compare it with expectations from cosmological simulations, and the LMC itself.

Meanwhile, both Sculptor A and B are extremely faint, quenched dwarf galaxies in isolated environments ($\gtrsim$3.7 $r_{vir, NGC~300}$ and $\gtrsim$2.3 $r_{vir, NGC~300}$ from NGC~300, respectively).  It is possible that these systems are extreme backsplash systems \cite[see discussion in][]{Santos23} and that their lack of gas and recent star formation is due to ram pressure or tidal stripping during a close passage with a larger galaxy. Recent simulations have suggested that some backsplash systems may be evident out to $\sim$3--4 $r_{vir}$ \citep[e.g.][]{Benavides21}, although similar simulations should be run for Magellanic Cloud-mass systems.
To investigate this further, a velocity measurement of these dwarfs is necessary, along with further distance constraints.

If one or both of Sculptor A \& B are not backsplash systems, then other mechanisms must be invoked to explain the lack of star formation and neutral gas.  Isolated field dwarfs with $M_*$$>$10$^7$ $M_{\odot}$ are almost all star-forming systems \citep{Geha12}, but below this threshold the quenched fraction is $\gtrsim$20\%  both in observations \citep{Slater14} and simulations \citep[e.g.][]{Christensen24}.  Without an interaction with a massive galaxy, Sculptor A or B may have been quenched by reionization and/or internal feedback (i.e. star formation or supernova feedback), which is seen in simulations of field dwarf galaxies \citep[e.g.][]{Jeon17,Rey20,Applebaum21}.  Observations of the ultra-faint dwarfs of the Milky Way lend support to this picture \citep{Weisz14,Brown14,Sacchi21,McQuinn24}, although the confounding effects of ram pressure and tidal stripping complicate the interpretation. Confirming this picture for field dwarfs would be a strong verification of galaxy formation models at the lowest mass scale.

Future \hi, spectroscopic and space-based data are needed to further understand the Sculptor dwarfs A, B and C.  A faint, extended gas reservoir, possibly offset from the optical emission, may signal the re-accretion of gas after reionization \citep[e.g.][]{Rey22}, and deep James Webb Space Telescope observations down to the oldest main sequence turnoff would reveal the earliest epoch of star formation \citep[e.g.][]{Weisz19}. Bulk velocity information will shed light on the orbital and interaction history of the dwarfs as well.  

Many more faint and ultra-faint dwarf galaxies are predicted at the edges of the Local Group and in nearby, low-density environments \citep{Tollerud18}, but initial efforts to find them have not always been successful \citep[e.g.][]{Tollerud15,Sand_UCHVC}.  Several upcoming programs such at Euclid \citep{euclid}, the Roman Space Telescope \citep{roman}, and the Rubin Observatory Legacy Survey of Space and Time \citep{lsst} are sure to find many more examples in the years ahead \citep[e.g.][]{Rodriguez19,Mutlu_sims}, which will provide demographic properties across environments.

\acknowledgments

Work on nearby galaxies by DJS and the Arizona team acknowledges support from NSF grant AST-2205863. Research by DC is supported by NSF grant AST-1814208. KS acknowledges support from the Natural Sciences and Engineering Research Council of Canada (NSERC).  J.E.A.\ and C.E.M-V are supported by the international Gemini Observatory, a program of NSF's NOIRLab, which is managed by the Association of Universities for Research in Astronomy (AURA) under a cooperative agreement with the National Science Foundation, on behalf of the Gemini partnership of Argentina, Brazil, Canada, Chile, the Republic of Korea, and the United States of America.

Based on observations obtained at the international Gemini Observatory under program GS-2022B-FT-103.  Gemini Observatory is a program of NSF’s NOIRLab, which is managed by the Association of Universities for Research in Astronomy (AURA) under a cooperative agreement with the National Science Foundation on behalf of the Gemini Observatory partnership: the National Science Foundation (United States), National Research Council (Canada), Agencia Nacional de Investigaci\'{o}n y Desarrollo (Chile), Ministerio de Ciencia, Tecnolog\'{i}a e Innovaci\'{o}n (Argentina), Minist\'{e}rio da Ci\^{e}ncia, Tecnologia, Inova\c{c}\~{o}es e Comunica\c{c}\~{o}es (Brazil), and Korea Astronomy and Space Science Institute (Republic of Korea).

The work used images from the Dark Energy Camera
Legacy Survey (DECaLS; Proposal ID 2014B-0404; PIs:
David Schlegel and Arjun Dey). Full acknowledgment
at \url{https://www.legacysurvey.org/acknowledgment/}. 

This research is based on observations made with the {\it Galaxy Evolution Explorer}, obtained from the MAST data archive at the Space Telescope Science Institute, which is operated by the Association of Universities for Research in Astronomy, Inc., under NASA contract NAS 5–26555.

This work was performed in part at the Aspen Center for Physics, which is supported by National Science Foundation grant PHY-2210452.

\vspace{5mm}
\facilities{Gemini:South (GMOS), {\it GALEX}}

\software{  astropy \citep{2013A&A...558A..33A,astropy}, {\sc astrometry.net} \citep{astrometry}
The IDL Astronomy User's Library \citep{IDLforever}, {\sc daophot} \citep{Stetson87,Stetson94}, {\sc dragons} \citep{dragonsRNAAS_2023}, \cite[Version 3.1.0]{dragons3.1.0}, {\sc scamp} \citep{scamp}, {\sc swarp} \citep{swarp}
          }

\bibliography{biblio}

\begin{thebibliography}{}
\expandafter\ifx\csname natexlab\endcsname\relax\def\natexlab#1{#1}\fi
\providecommand{\url}[1]{\href{#1}{#1}}
\providecommand{\dodoi}[1]{doi:~\href{http://doi.org/#1}{\nolinkurl{#1}}}
\providecommand{\doeprint}[1]{\href{http://ascl.net/#1}{\nolinkurl{http://ascl.net/#1}}}
\providecommand{\doarXiv}[1]{\href{https://arxiv.org/abs/#1}{\nolinkurl{https://arxiv.org/abs/#1}}}

\bibitem[{{Abbott} {et~al.}(2021){Abbott}, {Adam{\'o}w}, {Aguena}, {Allam},
  {Amon}, {Annis}, {Avila}, {Bacon}, {Banerji}, {Bechtol}, {Becker},
  {Bernstein}, {Bertin}, {Bhargava}, {Bridle}, {Brooks}, {Burke}, {Carnero
  Rosell}, {Carrasco Kind}, {Carretero}, {Castander}, {Cawthon}, {Chang},
  {Choi}, {Conselice}, {Costanzi}, {Crocce}, {da Costa}, {Davis}, {De Vicente},
  {DeRose}, {Desai}, {Diehl}, {Dietrich}, {Drlica-Wagner}, {Eckert},
  {Elvin-Poole}, {Everett}, {Evrard}, {Ferrero}, {Fert{\'e}}, {Flaugher},
  {Fosalba}, {Friedel}, {Frieman}, {Garc{\'\i}a-Bellido}, {Gaztanaga},
  {Gelman}, {Gerdes}, {Giannantonio}, {Gill}, {Gruen}, {Gruendl}, {Gschwend},
  {Gutierrez}, {Hartley}, {Hinton}, {Hollowood}, {Honscheid}, {Huterer},
  {James}, {Jeltema}, {Johnson}, {Kent}, {Kron}, {Kuehn}, {Kuropatkin},
  {Lahav}, {Li}, {Lidman}, {Lin}, {MacCrann}, {Maia}, {Manning}, {Maloney},
  {March}, {Marshall}, {Martini}, {Melchior}, {Menanteau}, {Miquel}, {Morgan},
  {Myles}, {Neilsen}, {Ogando}, {Palmese}, {Paz-Chinch{\'o}n}, {Petravick},
  {Pieres}, {Plazas}, {Pond}, {Rodriguez-Monroy}, {Romer}, {Roodman}, {Rykoff},
  {Sako}, {Sanchez}, {Santiago}, {Scarpine}, {Serrano}, {Sevilla-Noarbe},
  {Smith}, {Smith}, {Soares-Santos}, {Suchyta}, {Swanson}, {Tarle}, {Thomas},
  {To}, {Tremblay}, {Troxel}, {Tucker}, {Turner}, {Varga}, {Walker},
  {Wechsler}, {Weller}, {Wester}, {Wilkinson}, {Yanny}, {Zhang}, {Nikutta},
  {Fitzpatrick}, {Jacques}, {Scott}, {Olsen}, {Huang}, {Herrera}, {Juneau},
  {Nidever}, {Weaver}, {Adean}, {Correia}, {de Freitas}, {Freitas},
  {Singulani}, {Vila-Verde}, \& {Linea Science Server}}]{DES_DR2}
{Abbott}, T.~M.~C., {Adam{\'o}w}, M., {Aguena}, M., {et~al.} 2021, \apjs, 255,
  20, \dodoi{10.3847/1538-4365/ac00b3}

\bibitem[{{Ahvazi} {et~al.}(2024){Ahvazi}, {Benson}, {Sales}, {Nadler},
  {Weerasooriya}, {Du}, \& {Bovill}}]{Ahvazi24}
{Ahvazi}, N., {Benson}, A., {Sales}, L.~V., {et~al.} 2024, \mnras, 529, 3387,
  \dodoi{10.1093/mnras/stae761}

\bibitem[{{Akeson} {et~al.}(2019){Akeson}, {Armus}, {Bachelet}, {Bailey},
  {Bartusek}, {Bellini}, {Benford}, {Bennett}, {Bhattacharya}, {Bohlin},
  {Boyer}, {Bozza}, {Bryden}, {Calchi Novati}, {Carpenter}, {Casertano},
  {Choi}, {Content}, {Dayal}, {Dressler}, {Dor{\'e}}, {Fall}, {Fan}, {Fang},
  {Filippenko}, {Finkelstein}, {Foley}, {Furlanetto}, {Kalirai}, {Gaudi},
  {Gilbert}, {Girard}, {Grady}, {Greene}, {Guhathakurta}, {Heinrich},
  {Hemmati}, {Hendel}, {Henderson}, {Henning}, {Hirata}, {Ho}, {Huff},
  {Hutter}, {Jansen}, {Jha}, {Johnson}, {Jones}, {Kasdin}, {Kelly}, {Kirshner},
  {Koekemoer}, {Kruk}, {Lewis}, {Macintosh}, {Madau}, {Malhotra}, {Mandel},
  {Massara}, {Masters}, {McEnery}, {McQuinn}, {Melchior}, {Melton},
  {Mennesson}, {Peeples}, {Penny}, {Perlmutter}, {Pisani}, {Plazas}, {Poleski},
  {Postman}, {Ranc}, {Rauscher}, {Rest}, {Roberge}, {Robertson}, {Rodney},
  {Rhoads}, {Rhodes}, {Ryan}, {Sahu}, {Sand}, {Scolnic}, {Seth}, {Shvartzvald},
  {Siellez}, {Smith}, {Spergel}, {Stassun}, {Street}, {Strolger}, {Szalay},
  {Trauger}, {Troxel}, {Turnbull}, {van der Marel}, {von der Linden}, {Wang},
  {Weinberg}, {Williams}, {Windhorst}, {Wollack}, {Wu}, {Yee}, \&
  {Zimmerman}}]{roman}
{Akeson}, R., {Armus}, L., {Bachelet}, E., {et~al.} 2019, arXiv e-prints,
  arXiv:1902.05569, \dodoi{10.48550/arXiv.1902.05569}

\bibitem[{{Applebaum} {et~al.}(2021){Applebaum}, {Brooks}, {Christensen},
  {Munshi}, {Quinn}, {Shen}, \& {Tremmel}}]{Applebaum21}
{Applebaum}, E., {Brooks}, A.~M., {Christensen}, C.~R., {et~al.} 2021, \apj,
  906, 96, \dodoi{10.3847/1538-4357/abcafa}

\bibitem[{{Astropy Collaboration} {et~al.}(2013){Astropy Collaboration},
  {Robitaille}, {Tollerud}, {Greenfield}, {Droettboom}, {Bray}, {Aldcroft},
  {Davis}, {Ginsburg}, {Price-Whelan}, {Kerzendorf}, {Conley}, {Crighton},
  {Barbary}, {Muna}, {Ferguson}, {Grollier}, {Parikh}, {Nair}, {Unther},
  {Deil}, {Woillez}, {Conseil}, {Kramer}, {Turner}, {Singer}, {Fox}, {Weaver},
  {Zabalza}, {Edwards}, {Azalee Bostroem}, {Burke}, {Casey}, {Crawford},
  {Dencheva}, {Ely}, {Jenness}, {Labrie}, {Lim}, {Pierfederici}, {Pontzen},
  {Ptak}, {Refsdal}, {Servillat}, \& {Streicher}}]{2013A&A...558A..33A}
{Astropy Collaboration}, {Robitaille}, T.~P., {Tollerud}, E.~J., {et~al.} 2013,
  \aap, 558, A33, \dodoi{10.1051/0004-6361/201322068}

\bibitem[{{Barnes} {et~al.}(2001){Barnes}, {Staveley-Smith}, {de Blok},
  {Oosterloo}, {Stewart}, {Wright}, {Banks}, {Bhathal}, {Boyce}, {Calabretta},
  {Disney}, {Drinkwater}, {Ekers}, {Freeman}, {Gibson}, {Green}, {Haynes}, {te
  Lintel Hekkert}, {Henning}, {Jerjen}, {Juraszek}, {Kesteven}, {Kilborn},
  {Knezek}, {Koribalski}, {Kraan-Korteweg}, {Malin}, {Marquarding}, {Minchin},
  {Mould}, {Price}, {Putman}, {Ryder}, {Sadler}, {Schr{\"o}der}, {Stootman},
  {Webster}, {Wilson}, \& {Ye}}]{Barnes+2001}
{Barnes}, D.~G., {Staveley-Smith}, L., {de Blok}, W.~J.~G., {et~al.} 2001,
  \mnras, 322, 486, \dodoi{10.1046/j.1365-8711.2001.04102.x}

\bibitem[{{Battaglia} {et~al.}(2022){Battaglia}, {Taibi}, {Thomas}, \&
  {Fritz}}]{Battaglia22}
{Battaglia}, G., {Taibi}, S., {Thomas}, G.~F., \& {Fritz}, T.~K. 2022, \aap,
  657, A54, \dodoi{10.1051/0004-6361/202141528}

\bibitem[{{Benavides} {et~al.}(2021){Benavides}, {Sales}, {Abadi}, {Pillepich},
  {Nelson}, {Marinacci}, {Cooper}, {Pakmor}, {Torrey}, {Vogelsberger}, \&
  {Hernquist}}]{Benavides21}
{Benavides}, J.~A., {Sales}, L.~V., {Abadi}, M.~G., {et~al.} 2021, Nature
  Astronomy, 5, 1255, \dodoi{10.1038/s41550-021-01458-1}

\bibitem[{{Bennet} {et~al.}(2019){Bennet}, {Sand}, {Crnojevi{\'c}}, {Spekkens},
  {Karunakaran}, {Zaritsky}, \& {Mutlu-Pakdil}}]{Bennet19}
{Bennet}, P., {Sand}, D.~J., {Crnojevi{\'c}}, D., {et~al.} 2019, \apj, 885,
  153, \dodoi{10.3847/1538-4357/ab46ab}

\bibitem[{{Bennet} {et~al.}(2020){Bennet}, {Sand}, {Crnojevi{\'c}}, {Spekkens},
  {Karunakaran}, {Zaritsky}, \& {Mutlu-Pakdil}}]{Bennet20}
---. 2020, \apjl, 893, L9, \dodoi{10.3847/2041-8213/ab80c5}

\bibitem[{{Bennet} {et~al.}(2022){Bennet}, {Sand}, {Crnojevi{\'c}}, {Weisz},
  {Caldwell}, {Guhathakurta}, {Hargis}, {Karunakaran}, {Mutlu-Pakdil},
  {Olszewski}, {Salzer}, {Seth}, {Simon}, {Spekkens}, {Stark}, {Strader},
  {Tollerud}, {Toloba}, \& {Willman}}]{Bennet22}
---. 2022, \apj, 924, 98, \dodoi{10.3847/1538-4357/ac356c}

\bibitem[{{Benson} {et~al.}(2002){Benson}, {Frenk}, {Lacey}, {Baugh}, \&
  {Cole}}]{Benson02}
{Benson}, A.~J., {Frenk}, C.~S., {Lacey}, C.~G., {Baugh}, C.~M., \& {Cole}, S.
  2002, \mnras, 333, 177, \dodoi{10.1046/j.1365-8711.2002.05388.x}

\bibitem[{{Bertin}(2006)}]{scamp}
{Bertin}, E. 2006, in Astronomical Society of the Pacific Conference Series,
  Vol. 351, Astronomical Data Analysis Software and Systems XV, ed.
  C.~{Gabriel}, C.~{Arviset}, D.~{Ponz}, \& S.~{Enrique}, 112

\bibitem[{{Bertin}(2010)}]{swarp}
{Bertin}, E. 2010, {SWarp: Resampling and Co-adding FITS Images Together}.
\newblock \doeprint{1010.068}

\bibitem[{{Bianchi} {et~al.}(2017){Bianchi}, {Shiao}, \& {Thilker}}]{Bianchi17}
{Bianchi}, L., {Shiao}, B., \& {Thilker}, D. 2017, \apjs, 230, 24,
  \dodoi{10.3847/1538-4365/aa7053}

\bibitem[{{Brooks} {et~al.}(2013){Brooks}, {Kuhlen}, {Zolotov}, \&
  {Hooper}}]{Brooks13}
{Brooks}, A.~M., {Kuhlen}, M., {Zolotov}, A., \& {Hooper}, D. 2013, \apj, 765,
  22, \dodoi{10.1088/0004-637X/765/1/22}

\bibitem[{{Brown} {et~al.}(2014){Brown}, {Tumlinson}, {Geha}, {Simon},
  {Vargas}, {VandenBerg}, {Kirby}, {Kalirai}, {Avila}, {Gennaro}, {Ferguson},
  {Mu{\~n}oz}, {Guhathakurta}, \& {Renzini}}]{Brown14}
{Brown}, T.~M., {Tumlinson}, J., {Geha}, M., {et~al.} 2014, \apj, 796, 91,
  \dodoi{10.1088/0004-637X/796/2/91}

\bibitem[{{Buck} {et~al.}(2019){Buck}, {Macci{\`o}}, {Dutton}, {Obreja}, \&
  {Frings}}]{Buck19}
{Buck}, T., {Macci{\`o}}, A.~V., {Dutton}, A.~A., {Obreja}, A., \& {Frings}, J.
  2019, \mnras, 483, 1314, \dodoi{10.1093/mnras/sty2913}

\bibitem[{{Bullock} \& {Boylan-Kolchin}(2017)}]{Bullock17}
{Bullock}, J.~S., \& {Boylan-Kolchin}, M. 2017, \araa, 55, 343,
  \dodoi{10.1146/annurev-astro-091916-055313}

\bibitem[{{Bullock} {et~al.}(2000){Bullock}, {Kravtsov}, \&
  {Weinberg}}]{Bullock00}
{Bullock}, J.~S., {Kravtsov}, A.~V., \& {Weinberg}, D.~H. 2000, \apj, 539, 517,
  \dodoi{10.1086/309279}

\bibitem[{{Cannon} {et~al.}(2011){Cannon}, {Giovanelli}, {Haynes},
  {Janowiecki}, {Parker}, {Salzer}, {Adams}, {Engstrom}, {Huang}, {McQuinn},
  {Ott}, {Saintonge}, {Skillman}, {Allan}, {Erny}, {Fliss}, \&
  {Smith}}]{Cannon11}
{Cannon}, J.~M., {Giovanelli}, R., {Haynes}, M.~P., {et~al.} 2011, \apjl, 739,
  L22, \dodoi{10.1088/2041-8205/739/1/L22}

\bibitem[{{Carlin} {et~al.}(2016){Carlin}, {Sand}, {Price}, {Willman},
  {Karunakaran}, {Spekkens}, {Bell}, {Brodie}, {Crnojevi{\'c}}, {Forbes},
  {Hargis}, {Kirby}, {Lupton}, {Peter}, {Romanowsky}, \& {Strader}}]{Carlin16}
{Carlin}, J.~L., {Sand}, D.~J., {Price}, P., {et~al.} 2016, \apjl, 828, L5,
  \dodoi{10.3847/2041-8205/828/1/L5}

\bibitem[{{Carlin} {et~al.}(2021){Carlin}, {Mutlu-Pakdil}, {Crnojevi{\'c}},
  {Garling}, {Karunakaran}, {Peter}, {Tollerud}, {Forbes}, {Hargis}, {Lim},
  {Romanowsky}, {Sand}, {Spekkens}, \& {Strader}}]{Carlin21}
{Carlin}, J.~L., {Mutlu-Pakdil}, B., {Crnojevi{\'c}}, D., {et~al.} 2021, \apj,
  909, 211, \dodoi{10.3847/1538-4357/abe040}

\bibitem[{{Carlin} {et~al.}(2024){Carlin}, {Sand}, {Mutlu-Pakdil}, {Crnojevic},
  {Doliva-Dolinsky}, {Garling}, {Peter}, {Brodie}, {Forbes}, {Hargis},
  {Romanowsky}, {Spekkens}, {Strader}, \& {Willman}}]{Carlin24}
{Carlin}, J.~L., {Sand}, D.~J., {Mutlu-Pakdil}, B., {et~al.} 2024, arXiv
  e-prints, arXiv:2409.17437, \dodoi{10.48550/arXiv.2409.17437}

\bibitem[{{Carlsten} {et~al.}(2022){Carlsten}, {Greene}, {Beaton}, {Danieli},
  \& {Greco}}]{Carlsten22}
{Carlsten}, S.~G., {Greene}, J.~E., {Beaton}, R.~L., {Danieli}, S., \& {Greco},
  J.~P. 2022, \apj, 933, 47, \dodoi{10.3847/1538-4357/ac6fd7}

\bibitem[{{Casey} {et~al.}(2023){Casey}, {Greco}, {Peter}, \&
  {Davis}}]{Casey23}
{Casey}, K.~J., {Greco}, J.~P., {Peter}, A. H.~G., \& {Davis}, A.~B. 2023,
  \mnras, 520, 4715, \dodoi{10.1093/mnras/stad352}

\bibitem[{{Cerny} {et~al.}(2023){Cerny}, {Mart{\'\i}nez-V{\'a}zquez},
  {Drlica-Wagner}, {Pace}, {Mutlu-Pakdil}, {Li}, {Riley}, {Crnojevi{\'c}},
  {Bom}, {Carballo-Bello}, {Carlin}, {Chiti}, {Choi}, {Collins},
  {Darragh-Ford}, {Ferguson}, {Geha}, {Mart{\'\i}nez-Delgado}, {Massana},
  {Mau}, {Medina}, {Mu{\~n}oz}, {Nadler}, {No{\"e}l}, {Olsen}, {Pieres},
  {Sakowska}, {Simon}, {Stringfellow}, {Tollerud}, {Vivas}, {Walker},
  {Wechsler}, \& {Delve Collaboration}}]{Cerny23}
{Cerny}, W., {Mart{\'\i}nez-V{\'a}zquez}, C.~E., {Drlica-Wagner}, A., {et~al.}
  2023, \apj, 953, 1, \dodoi{10.3847/1538-4357/acdd78}

\bibitem[{{Chiboucas} {et~al.}(2013){Chiboucas}, {Jacobs}, {Tully}, \&
  {Karachentsev}}]{Chiboucas13}
{Chiboucas}, K., {Jacobs}, B.~A., {Tully}, R.~B., \& {Karachentsev}, I.~D.
  2013, \aj, 146, 126, \dodoi{10.1088/0004-6256/146/5/126}

\bibitem[{{Christensen} {et~al.}(2024){Christensen}, {Brooks}, {Munshi},
  {Riggs}, {Van Nest}, {Akins}, {Quinn}, \& {Chamberland}}]{Christensen24}
{Christensen}, C.~R., {Brooks}, A.~M., {Munshi}, F., {et~al.} 2024, \apj, 961,
  236, \dodoi{10.3847/1538-4357/ad0c5a}

\bibitem[{{Crnojevi{\'c}} {et~al.}(2016{\natexlab{a}}){Crnojevi{\'c}}, {Sand},
  {Zaritsky}, {Spekkens}, {Willman}, \& {Hargis}}]{Crnojevic_erii}
{Crnojevi{\'c}}, D., {Sand}, D.~J., {Zaritsky}, D., {et~al.}
  2016{\natexlab{a}}, \apjl, 824, L14, \dodoi{10.3847/2041-8205/824/1/L14}

\bibitem[{{Crnojevi{\'c}} {et~al.}(2016{\natexlab{b}}){Crnojevi{\'c}}, {Sand},
  {Spekkens}, {Caldwell}, {Guhathakurta}, {McLeod}, {Seth}, {Simon}, {Strader},
  \& {Toloba}}]{Crnojevic16}
{Crnojevi{\'c}}, D., {Sand}, D.~J., {Spekkens}, K., {et~al.}
  2016{\natexlab{b}}, \apj, 823, 19, \dodoi{10.3847/0004-637X/823/1/19}

\bibitem[{{Crnojevi{\'c}} {et~al.}(2019){Crnojevi{\'c}}, {Sand}, {Bennet},
  {Pasetto}, {Spekkens}, {Caldwell}, {Guhathakurta}, {McLeod}, {Seth}, {Simon},
  {Strader}, \& {Toloba}}]{Crnojevic19}
{Crnojevi{\'c}}, D., {Sand}, D.~J., {Bennet}, P., {et~al.} 2019, \apj, 872, 80,
  \dodoi{10.3847/1538-4357/aafbe7}

\bibitem[{{Dalcanton} {et~al.}(2009){Dalcanton}, {Williams}, {Seth}, {Dolphin},
  {Holtzman}, {Rosema}, {Skillman}, {Cole}, {Girardi}, {Gogarten},
  {Karachentsev}, {Olsen}, {Weisz}, {Christensen}, {Freeman}, {Gilbert},
  {Gallart}, {Harris}, {Hodge}, {de Jong}, {Karachentseva}, {Mateo}, {Stetson},
  {Tavarez}, {Zaritsky}, {Governato}, \& {Quinn}}]{angst}
{Dalcanton}, J.~J., {Williams}, B.~F., {Seth}, A.~C., {et~al.} 2009, \apjs,
  183, 67, \dodoi{10.1088/0067-0049/183/1/67}

\bibitem[{{Davis} {et~al.}(2024){Davis}, {Garling}, {Nierenberg}, {Peter},
  {Sardone}, {Kochanek}, {Leroy}, {Casey}, {Pogge}, {Roberts}, {Sand}, \&
  {Greco}}]{Davis24}
{Davis}, A.~B., {Garling}, C.~T., {Nierenberg}, A.~M., {et~al.} 2024, arXiv
  e-prints, arXiv:2409.03999.
\newblock \doarXiv{2409.03999}

\bibitem[{{Dekel} \& {Silk}(1986)}]{Dekel86}
{Dekel}, A., \& {Silk}, J. 1986, \apj, 303, 39, \dodoi{10.1086/164050}

\bibitem[{{Dey} {et~al.}(2019){Dey}, {Schlegel}, {Lang}, {Blum}, {Burleigh},
  {Fan}, {Findlay}, {Finkbeiner}, {Herrera}, {Juneau}, {Landriau}, {Levi},
  {McGreer}, {Meisner}, {Myers}, {Moustakas}, {Nugent}, {Patej}, {Schlafly},
  {Walker}, {Valdes}, {Weaver}, {Y{\`e}che}, {Zou}, {Zhou}, {Abareshi},
  {Abbott}, {Abolfathi}, {Aguilera}, {Alam}, {Allen}, {Alvarez}, {Annis},
  {Ansarinejad}, {Aubert}, {Beechert}, {Bell}, {BenZvi}, {Beutler}, {Bielby},
  {Bolton}, {Brice{\~n}o}, {Buckley-Geer}, {Butler}, {Calamida}, {Carlberg},
  {Carter}, {Casas}, {Castander}, {Choi}, {Comparat}, {Cukanovaite}, {Delubac},
  {DeVries}, {Dey}, {Dhungana}, {Dickinson}, {Ding}, {Donaldson}, {Duan},
  {Duckworth}, {Eftekharzadeh}, {Eisenstein}, {Etourneau}, {Fagrelius},
  {Farihi}, {Fitzpatrick}, {Font-Ribera}, {Fulmer}, {G{\"a}nsicke},
  {Gaztanaga}, {George}, {Gerdes}, {Gontcho}, {Gorgoni}, {Green}, {Guy},
  {Harmer}, {Hernandez}, {Honscheid}, {Huang}, {James}, {Jannuzi}, {Jiang},
  {Joyce}, {Karcher}, {Karkar}, {Kehoe}, {Kneib}, {Kueter-Young}, {Lan},
  {Lauer}, {Le Guillou}, {Le Van Suu}, {Lee}, {Lesser}, {Perreault Levasseur},
  {Li}, {Mann}, {Marshall}, {Mart{\'\i}nez-V{\'a}zquez}, {Martini}, {du Mas des
  Bourboux}, {McManus}, {Meier}, {M{\'e}nard}, {Metcalfe},
  {Mu{\~n}oz-Guti{\'e}rrez}, {Najita}, {Napier}, {Narayan}, {Newman}, {Nie},
  {Nord}, {Norman}, {Olsen}, {Paat}, {Palanque-Delabrouille}, {Peng},
  {Poppett}, {Poremba}, {Prakash}, {Rabinowitz}, {Raichoor}, {Rezaie},
  {Robertson}, {Roe}, {Ross}, {Ross}, {Rudnick}, {Safonova}, {Saha},
  {S{\'a}nchez}, {Savary}, {Schweiker}, {Scott}, {Seo}, {Shan}, {Silva},
  {Slepian}, {Soto}, {Sprayberry}, {Staten}, {Stillman}, {Stupak}, {Summers},
  {Sien Tie}, {Tirado}, {Vargas-Maga{\~n}a}, {Vivas}, {Wechsler}, {Williams},
  {Yang}, {Yang}, {Yapici}, {Zaritsky}, {Zenteno}, {Zhang}, {Zhang}, {Zhou}, \&
  {Zhou}}]{Dey19}
{Dey}, A., {Schlegel}, D.~J., {Lang}, D., {et~al.} 2019, \aj, 157, 168,
  \dodoi{10.3847/1538-3881/ab089d}

\bibitem[{{Diemer} \& {Kravtsov}(2015)}]{Diemer15}
{Diemer}, B., \& {Kravtsov}, A.~V. 2015, \apj, 799, 108,
  \dodoi{10.1088/0004-637X/799/1/108}

\bibitem[{{Doliva-Dolinsky} {et~al.}(2022){Doliva-Dolinsky}, {Martin},
  {Thomas}, {Ferguson}, {Ibata}, {Lewis}, {Mackey}, {McConnachie}, \&
  {Yuan}}]{DD22}
{Doliva-Dolinsky}, A., {Martin}, N.~F., {Thomas}, G.~F., {et~al.} 2022, \apj,
  933, 135, \dodoi{10.3847/1538-4357/ac6fd5}

\bibitem[{{Doliva-Dolinsky} {et~al.}(2023){Doliva-Dolinsky}, {Martin}, {Yuan},
  {Savino}, {Weisz}, {Ferguson}, {Ibata}, {Kim}, {Lewis}, {McConnachie}, \&
  {Thomas}}]{DD23}
{Doliva-Dolinsky}, A., {Martin}, N.~F., {Yuan}, Z., {et~al.} 2023, \apj, 952,
  72, \dodoi{10.3847/1538-4357/acdcf6}

\bibitem[{{Dotter} {et~al.}(2008){Dotter}, {Chaboyer}, {Jevremovi{\'c}},
  {Kostov}, {Baron}, \& {Ferguson}}]{Dotter08}
{Dotter}, A., {Chaboyer}, B., {Jevremovi{\'c}}, D., {et~al.} 2008, \apjs, 178,
  89, \dodoi{10.1086/589654}

\bibitem[{{Drlica-Wagner} {et~al.}(2020){Drlica-Wagner}, {Bechtol}, {Mau},
  {McNanna}, {Nadler}, {Pace}, {Li}, {Pieres}, {Rozo}, {Simon}, {Walker},
  {Wechsler}, {Abbott}, {Allam}, {Annis}, {Bertin}, {Brooks}, {Burke},
  {Rosell}, {Carrasco Kind}, {Carretero}, {Costanzi}, {da Costa}, {De Vicente},
  {Desai}, {Diehl}, {Doel}, {Eifler}, {Everett}, {Flaugher}, {Frieman},
  {Garc{\'\i}a-Bellido}, {Gaztanaga}, {Gruen}, {Gruendl}, {Gschwend},
  {Gutierrez}, {Honscheid}, {James}, {Krause}, {Kuehn}, {Kuropatkin}, {Lahav},
  {Maia}, {Marshall}, {Melchior}, {Menanteau}, {Miquel}, {Palmese}, {Plazas},
  {Sanchez}, {Scarpine}, {Schubnell}, {Serrano}, {Sevilla-Noarbe}, {Smith},
  {Suchyta}, {Tarle}, \& {DES Collaboration}}]{Drlica20}
{Drlica-Wagner}, A., {Bechtol}, K., {Mau}, S., {et~al.} 2020, \apj, 893, 47,
  \dodoi{10.3847/1538-4357/ab7eb9}

\bibitem[{{Drlica-Wagner} {et~al.}(2021){Drlica-Wagner}, {Carlin}, {Nidever},
  {Ferguson}, {Kuropatkin}, {Adam{\'o}w}, {Cerny}, {Choi}, {Esteves},
  {Mart{\'\i}nez-V{\'a}zquez}, {Mau}, {Miller}, {Mutlu-Pakdil}, {Neilsen},
  {Olsen}, {Pace}, {Riley}, {Sakowska}, {Sand}, {Santana-Silva}, {Tollerud},
  {Tucker}, {Vivas}, {Zaborowski}, {Zenteno}, {Abbott}, {Allam}, {Bechtol},
  {Bell}, {Bell}, {Bilaji}, {Bom}, {Carballo-Bello}, {Crnojevi{\'c}}, {Cioni},
  {Diaz-Ocampo}, {de Boer}, {Erkal}, {Gruendl}, {Hernandez-Lang}, {Hughes},
  {James}, {Johnson}, {Li}, {Mao}, {Mart{\'\i}nez-Delgado}, {Massana},
  {McNanna}, {Morgan}, {Nadler}, {No{\"e}l}, {Palmese}, {Peter}, {Rykoff},
  {S{\'a}nchez}, {Shipp}, {Simon}, {Smercina}, {Soares-Santos}, {Stringfellow},
  {Tavangar}, {van der Marel}, {Walker}, {Wechsler}, {Wu}, {Yanny},
  {Fitzpatrick}, {Huang}, {Jacques}, {Nikutta}, {Scott}, \& {Astro Data
  Lab}}]{DELVEdr1}
{Drlica-Wagner}, A., {Carlin}, J.~L., {Nidever}, D.~L., {et~al.} 2021, \apjs,
  256, 2, \dodoi{10.3847/1538-4365/ac079d}

\bibitem[{{Drlica-Wagner} {et~al.}(2022){Drlica-Wagner}, {Ferguson},
  {Adam{\'o}w}, {Aguena}, {Allam}, {Andrade-Oliveira}, {Bacon}, {Bechtol},
  {Bell}, {Bertin}, {Bilaji}, {Bocquet}, {Bom}, {Brooks}, {Burke},
  {Carballo-Bello}, {Carlin}, {Carnero Rosell}, {Carrasco Kind}, {Carretero},
  {Castander}, {Cerny}, {Chang}, {Choi}, {Conselice}, {Costanzi},
  {Crnojevi{\'c}}, {da Costa}, {de Vicente}, {Desai}, {Esteves}, {Everett},
  {Ferrero}, {Fitzpatrick}, {Flaugher}, {Friedel}, {Frieman},
  {Garc{\'\i}a-Bellido}, {Gatti}, {Gaztanaga}, {Gerdes}, {Gruen}, {Gruendl},
  {Gschwend}, {Hartley}, {Hernandez-Lang}, {Hinton}, {Hollowood}, {Honscheid},
  {Hughes}, {Jacques}, {James}, {Johnson}, {Kuehn}, {Kuropatkin}, {Lahav},
  {Li}, {Lidman}, {Lin}, {March}, {Marshall}, {Mart{\'\i}nez-Delgado},
  {Mart{\'\i}nez-V{\'a}zquez}, {Massana}, {Mau}, {McNanna}, {Melchior},
  {Menanteau}, {Miller}, {Miquel}, {Mohr}, {Morgan}, {Mutlu-Pakdil},
  {Mu{\~n}oz}, {Neilsen}, {Nidever}, {Nikutta}, {Nilo Castellon}, {No{\"e}l},
  {Ogando}, {Olsen}, {Pace}, {Palmese}, {Paz-Chinch{\'o}n}, {Pereira},
  {Pieres}, {Plazas Malag{\'o}n}, {Prat}, {Riley}, {Rodriguez-Monroy}, {Romer},
  {Roodman}, {Sako}, {Sakowska}, {Sanchez}, {S{\'a}nchez}, {Sand},
  {Santana-Silva}, {Santiago}, {Schubnell}, {Serrano}, {Sevilla-Noarbe},
  {Simon}, {Smith}, {Soares-Santos}, {Stringfellow}, {Suchyta}, {Suson}, {Tan},
  {Tarle}, {Tavangar}, {Thomas}, {To}, {Tollerud}, {Troxel}, {Tucker}, {Varga},
  {Vivas}, {Walker}, {Weller}, {Wilkinson}, {Wu}, {Yanny}, {Zaborowski},
  {Zenteno}, {Delve Collaboration}, {Des Collaboration}, \& {Astro Data
  Lab}}]{DELVEdr2}
{Drlica-Wagner}, A., {Ferguson}, P.~S., {Adam{\'o}w}, M., {et~al.} 2022, \apjs,
  261, 38, \dodoi{10.3847/1538-4365/ac78eb}

\bibitem[{{El-Badry} {et~al.}(2018){El-Badry}, {Quataert}, {Wetzel}, {Hopkins},
  {Weisz}, {Chan}, {Fitts}, {Boylan-Kolchin}, {Kere{\v{s}}},
  {Faucher-Gigu{\`e}re}, \& {Garrison-Kimmel}}]{Elbadry18}
{El-Badry}, K., {Quataert}, E., {Wetzel}, A., {et~al.} 2018, \mnras, 473, 1930,
  \dodoi{10.1093/mnras/stx2482}

\bibitem[{{Engler} {et~al.}(2021){Engler}, {Pillepich}, {Pasquali}, {Nelson},
  {Rodriguez-Gomez}, {Chua}, {Grebel}, {Springel}, {Marinacci}, {Weinberger},
  {Vogelsberger}, \& {Hernquist}}]{Engler21}
{Engler}, C., {Pillepich}, A., {Pasquali}, A., {et~al.} 2021, arXiv e-prints,
  arXiv:2101.12215.
\newblock \doarXiv{2101.12215}

\bibitem[{{Euclid Collaboration} {et~al.}(2024){Euclid Collaboration},
  {Mellier}, {Abdurro'uf}, {Acevedo Barroso}, {Ach{\'u}carro}, {Adamek},
  {Adam}, {Addison}, {Aghanim}, {Aguena}, {Ajani}, {Akrami}, {Al-Bahlawan},
  {Alavi}, {Albuquerque}, {Alestas}, {Alguero}, {Allaoui}, {Allen}, {Allevato},
  {Alonso-Tetilla}, {Altieri}, {Alvarez-Candal}, {Amara}, {Amendola}, {Amiaux},
  {Andika}, {Andreon}, {Andrews}, {Angora}, {Angulo}, {Annibali}, {Anselmi},
  {Anselmi}, {Arcari}, {Archidiacono}, {Aric{\`o}}, {Arnaud}, {Arnouts},
  {Asgari}, {Asorey}, {Atayde}, {Atek}, {Atrio-Barandela}, {Aubert}, {Aubourg},
  {Auphan}, {Auricchio}, {Aussel}, {Aussel}, {Avelino}, {Avgoustidis}, {Avila},
  {Awan}, {Azzollini}, {Baccigalupi}, {Bachelet}, {Bacon}, {Baes}, {Bagley},
  {Bahr-Kalus}, {Balaguera-Antolinez}, {Balbinot}, {Balcells}, {Baldi},
  {Baldry}, {Balestra}, {Ballardini}, {Ballester}, {Balogh}, {Ba{\~n}ados},
  {Barbier}, {Bardelli}, {Barreiro}, {Barriere}, {Barros}, {Barthelemy},
  {Bartolo}, {Basset}, {Battaglia}, {Battisti}, {Baugh}, {Baumont},
  {Bazzanini}, {Beaulieu}, {Beckmann}, {Belikov}, {Bel}, {Bellagamba}, {Bella},
  {Bellini}, {Benabed}, {Bender}, {Benevento}, {Bennett}, {Benson},
  {Bergamini}, {Bermejo-Climent}, {Bernardeau}, {Bertacca}, {Berthe},
  {Berthier}, {Bethermin}, {Beutler}, {Bevillon}, {Bhargava}, {Bhatawdekar},
  {Bisigello}, {Biviano}, {Blake}, {Blanchard}, {Blazek}, {Blot}, {Bosco},
  {Bodendorf}, {Boenke}, {B{\"o}hringer}, {Bolzonella}, {Bonchi}, {Bonici},
  {Bonino}, {Bonino}, {Bonvin}, {Bon}, {Booth}, {Borgani}, {Borlaff},
  {Borsato}, {Bosco}, {Bose}, {Botticella}, {Boucaud}, {Bouche}, {Boucher},
  {Boutigny}, {Bouvard}, {Bouy}, {Bowler}, {Bozza}, {Bozzo}, {Branchini},
  {Brau-Nogue}, {Brekke}, {Bremer}, {Brescia}, {Breton}, {Brinchmann},
  {Brinckmann}, {Brockley-Blatt}, {Brodwin}, {Brouard}, {Brown}, {Bruton},
  {Bucko}, {Buddelmeijer}, {Buenadicha}, {Buitrago}, {Burger}, {Burigana},
  {Busillo}, {Busonero}, {Cabanac}, {Cabayol-Garcia}, {Cagliari}, {Caillat},
  {Caillat}, {Calabrese}, {Calabro}, {Calderone}, {Calura}, {Camacho Quevedo},
  {Camera}, {Campos}, {Canas-Herrera}, {Candini}, {Cantiello}, {Capobianco},
  {Cappellaro}, {Cappelluti}, {Cappi}, {Caputi}, {Cara}, {Carbone}, {Cardone},
  {Carella}, {Carlberg}, {Carle}, {Carminati}, {Caro}, {Carrasco}, {Carretero},
  {Carrilho}, {Carron Duque}, {Carry}, {Carvalho}, {Carvalho}, {Casas},
  {Casas}, {Casenove}, {Casey}, {Cassata}, {Castander}, {Castelao},
  {Castellano}, {Castiblanco}, {Castignani}, {Castro}, {Cavet}, {Cavuoti},
  {Chabaud}, {Chambers}, {Charles}, {Charlot}, {Chartab}, {Chary}, {Chaumeil},
  {Cho}, {Chon}, {Ciancetta}, {Ciliegi}, {Cimatti}, {Cimino}, {Cioni},
  {Claydon}, {Cleland}, {Cl{\'e}ment}, {Clements}, {Clerc}, {Clesse}, {Codis},
  {Cogato}, {Colbert}, {Cole}, {Coles}, {Collett}, {Collins}, {Colodro-Conde},
  {Colombo}, {Combes}, {Conforti}, {Congedo}, {Conseil}, {Conselice},
  {Contarini}, {Contini}, {Conversi}, {Cooray}, {Copin}, {Corasaniti},
  {Corcho-Caballero}, {Corcione}, {Cordes}, {Corpace}, {Correnti}, {Costanzi},
  {Costille}, {Courbin}, {Courcoult Mifsud}, {Courtois}, {Cousinou}, {Covone},
  {Cowell}, {Cragg}, {Cresci}, {Cristiani}, {Crocce}, {Cropper}, {E Crouzet},
  {Csizi}, {Cuby}, {Cucchetti}, {Cucciati}, {Cuillandre}, {Cunha}, {Cuozzo},
  {Daddi}, {D'Addona}, {Dafonte}, {Dagoneau}, {Dalessandro}, {Dalton},
  {D'Amico}, {Dannerbauer}, {Danto}, {Das}, {Da Silva}, {da Silva}, {Daste},
  {Davies}, {Davini}, {de Boer}, {Decarli}, {De Caro}, {Degaudenzi}, {Degni},
  {de Jong}, {de la Bella}, {de la Torre}, {Delhaise}, {Delley}, {Delucchi},
  {De Lucia}, {Denniston}, {De Paolis}, {De Petris}, {Derosa}, {Desai},
  {Desjacques}, {Despali}, {Desprez}, {De Vicente-Albendea}, {Deville}, {Dias},
  {D{\'\i}az-S{\'a}nchez}, {Diaz}, {Di Domizio}, {Diego}, {Di Ferdinando}, {Di
  Giorgio}, {Dimauro}, {Dinis}, {Dolag}, {Dolding}, {Dole}, {Dom{\'\i}nguez
  S{\'a}nchez}, {Dor{\'e}}, {Dournac}, {Douspis}, {Dreihahn}, {Droge}, {Dryer},
  {Dubath}, {Duc}, {Ducret}, {Duffy}, {Dufresne}, {Duncan}, {Dupac}, {Duret},
  {Durrer}, {Durret}, {Dusini}, {Ealet}, {Eggemeier}, {Eisenhardt}, {Elbaz},
  {Elkhashab}, {Ellien}, {Endicott}, {Enia}, {Erben}, {Escartin Vigo},
  {Escoffier}, {Escudero Sanz}, {Essert}, {Ettori}, {Ezziati}, {Fabbian},
  {Fabricius}, {Fang}, {Farina}, {Farina}, {Farinelli}, {Farrens}, {Faustini},
  {Feltre}, {Ferguson}, {Ferrando}, {Ferrari}, {Ferr{\'e}-Mateu}, {Ferreira},
  {Ferreras}, {Ferrero}, {Ferriol}, {Ferruit}, {Filleul}, {Finelli},
  {Finkelstein}, {Finoguenov}, {Fiorini}, {Flentge}, {Focardi}, {Fonseca},
  {Fontana}, {Fontanot}, {Fornari}, {Fosalba}, {Fossati}, {Fotopoulou},
  {Fouchez}, {Fourmanoit}, {Frailis}, {Fraix-Burnet}, {Franceschi}, {Franco},
  {Franzetti}, {Freihoefer}, {Frittoli}, {Frugier}, {Frusciante}, {Fumagalli},
  {Fumagalli}, {Fumana}, {Fu}, {Gabarra}, {Galeotta}, {Galluccio}, {Ganga},
  {Gao}, {Garc{\'\i}a-Bellido}, {Garcia}, {Gardner}, {Garilli},
  {Gaspar-Venancio}, {Gasparetto}, {Gautard}, {Gavazzi}, {Gaztanaga},
  {Genolet}, {Genova Santos}, {Gentile}, {George}, {Ghaffari}, {Giacomini},
  {Gianotti}, {Gibb}, {Gillard}, {Gillis}, {Ginolfi}, {Giocoli}, {Girardi},
  {Giri}, {Goh}, {G{\'o}mez-Alvarez}, {Gonzalez}, {Gonzalez}, {Gonzalez},
  {Gouyou Beauchamps}, {Gozaliasl}, {Gracia-Carpio}, {Grandis}, {Granett},
  {Granvik}, {Grazian}, {Gregorio}, {Grenet}, {Grillo}, {Grupp}, {Gruppioni},
  {Gruppuso}, {Guerbuez}, {Guerrini}, {Guidi}, {Guillard}, {Gutierrez},
  {Guttridge}, {Guzzo}, {Gwyn}, {Haapala}, {Haase}, {Haddow}, {Hailey}, {Hall},
  {Hall}, {Hamaus}, {Haridasu}, {Harnois-D{\'e}raps}, {Harper}, {Hartley},
  {Hasinger}, {Hassani}, {Hatch}, {Haugan}, {H{\"a}u{\ss}ler}, {Heavens},
  {Heisenberg}, {Helmi}, {Helou}, {Hemmati}, {Henares}, {Herent},
  {Hern{\'a}ndez-Monteagudo}, {Heuberger}, {Hewett}, {Heydenreich},
  {Hildebrandt}, {Hirschmann}, {Hjorth}, {Hoar}, {Hoekstra}, {Holland},
  {Holliman}, {Holmes}, {Hook}, {Horeau}, {Hormuth}, {Hornstrup}, {Hosseini},
  {Hu}, {Hudelot}, {Hudson}, {Huertas-Company}, {Huff}, {Hughes}, {Humphrey},
  {Hunt}, {Huynh}, {Ibata}, {Ichikawa}, {Iglesias-Groth}, {Ilbert}, {Ili{\'c}},
  {Ingoglia}, {Iodice}, {Israel}, {Israelsson}, {Izzo}, {Jablonka}, {Jackson},
  {Jacobson}, {Jafariyazani}, {Jahnke}, {Jansen}, {Jarvis}, {Jasche}, {Jauzac},
  {Jeffrey}, {Jhabvala}, {Jimenez-Teja}, {Jimenez Mu{\~n}oz}, {Joachimi},
  {Johansson}, {Joudaki}, {Jullo}, {Kajava}, {Kang}, {Kannawadi}, {Kansal},
  {Karagiannis}, {K{\"a}rcher}, {Kashlinsky}, {Kazandjian}, {Keck},
  {Keih{\"a}nen}, {Kerins}, {Kermiche}, {Khalil}, {Kiessling}, {Kiiveri},
  {Kilbinger}, {Kim}, {King}, {Kirkpatrick}, {Kitching}, {Kluge}, {Knabenhans},
  {Knapen}, {Knebe}, {Kneib}, {Kohley}, {Koopmans}, {Koskinen}, {Koulouridis},
  {Kou}, {Kov{\'a}cs}, {Kova\{{\v{c}}\}i{\'c}}, {Kowalczyk}, {Koyama},
  {Kraljic}, {Krause}, {Kruk}, {Kubik}, {Kuchner}, {Kuijken}, {K{\"u}mmel},
  {Kunz}, {Kurki-Suonio}, {Lacasa}, {Lacey}, {La Franca}, {Lagarde}, {Lahav},
  {Laigle}, {La Marca}, {La Marle}, {Lamine}, {Lam}, {Lan{\c{c}}on}, {Landt},
  {Langer}, {Lapi}, {Larcheveque}, {Larsen}, {Lattanzi}, {Laudisio}, {Laugier},
  {Laureijs}, {Lavaux}, {Lawrenson}, {Lazanu}, {Lazeyras}, {Le Boulc'h}, {Le
  Brun}, {Le Brun}, {Leclercq}, {Lee}, {Le Graet}, {Legrand}, {Leirvik}, {Le
  Jeune}, {Lembo}, {Le Mignant}, {Lepinzan}, {Lepori}, {Lesci}, {Lesgourgues},
  {Leuzzi}, {Levi}, {Liaudat}, {Libet}, {Liebing}, {Ligori}, {Lilje}, {Lin},
  {Linde}, {Linder}, {Lindholm}, {Linke}, {Li}, {Liu}, {Lloro}, {Lobo},
  {Lodieu}, {Lombardi}, {Lombriser}, {Lonare}, {Longo}, {L{\'o}pez-Caniego},
  {Lopez Lopez}, {Alvarez}, {Loureiro}, {Loveday}, {Lusso}, {Macias-Perez},
  {Maciaszek}, {Magliocchetti}, {Magnard}, {Magnier}, {Magro}, {Mahler},
  {Mainetti}, {Maino}, {Maiorano}, {Maiorano}, {Malavasi}, {Mamon}, {Mancini},
  {Mandelbaum}, {Manera}, {Manj{\'o}n-Garc{\'\i}a}, {Mannucci}, {Mansutti},
  {Manteiga Outeiro}, {Maoli}, {Maraston}, {Marcin}, {Marcos-Arenal},
  {Margalef-Bentabol}, {Marggraf}, {Marinucci}, {Marinucci}, {Markovic},
  {Marleau}, {Marpaud}, {Martignac}, {Mart{\'\i}n-Fleitas}, {Martin-Moruno},
  {Martin}, {Martinelli}, {Martinet}, {Martin}, {Martins}, {Marulli},
  {Massari}, {Massey}, {Masters}, {Matarrese}, {Matsuoka}, {Matthew},
  {Maughan}, {Mauri}, {Maurin}, {Maurogordato}, {McCarthy}, {McConnachie},
  {McCracken}, {McDonald}, {McEwen}, {McPartland}, {Medinaceli}, {Mehta},
  {Mei}, {Melchior}, {Melin}, {M{\'e}nard}, {Mendes}, {Mendez-Abreu},
  {Meneghetti}, {Mercurio}, {Merlin}, {Metcalf}, {Meylan}, {Migliaccio},
  {Mignoli}, {Miller}, {Miluzio}, {Milvang-Jensen}, {Mimoso}, {Miquel},
  {Miyatake}, {Mobasher}, {Mohr}, {Monaco}, {Mongui{\'o}}, {Montoro}, {Mora},
  {Moradinezhad Dizgah}, {Moresco}, {Moretti}, {Morgante}, {Morisset},
  {Moriya}, {Morris}, {Mortlock}, {Moscardini}, {Mota}, {Moustakas}, {Moutard},
  {M{\"u}ller}, {Munari}, {Murphree}, {Murray}, {Murray}, {Musi}, {Nadathur},
  {Nagam}, {Nagao}, {Naidoo}, {Nakajima}, {Nally}, {Natoli}, {Navarro-Alsina},
  {Navarro Girones}, {Neissner}, {Nersesian}, {Nesseris}, {Nguyen-Kim},
  {Nicastro}, {Nichol}, {Nielbock}, {Niemi}, {Nieto}, {Nilsson}, {Noller},
  {Norberg}, {Nourizonoz}, {Ntelis}, {Nucita}, {Nugent}, {Nunes}, {Nutma},
  {Ocampo}, {Odier}, {Oesch}, {Oguri}, {Magalhaes Oliveira}, {Onoue},
  {Oosterbroek}, {Oppizzi}, {Ordenovic}, {Osato}, {Pacaud}, {Pace}, {Padilla},
  {Paech}, {Pagano}, {Page}, {Palazzi}, {Paltani}, {Pamuk}, {Pandolfi},
  {Paoletti}, {Paolillo}, {Papaderos}, {Pardede}, {Parimbelli}, {Parmar},
  {Partmann}, {Pasian}, {Passalacqua}, {Paterson}, {Patrizii}, {Pattison},
  {Paulino-Afonso}, {Paviot}, {Peacock}, {Pearce}, {Pedersen}, {Peel},
  {Peletier}, {Pellejero Ibanez}, {Pello}, {Penny}, {Percival},
  {Perez-Garrido}, {Perotto}, {Pettorino}, {Pezzotta}, {Pezzuto}, {Philippon},
  {Piersanti}, {Pietroni}, {Piga}, {Pilo}, {Pires}, {Pisani}, {Pizzella},
  {Pizzuti}, {Plana}, {Polenta}, {Pollack}, {Poncet}, {P{\"o}ntinen}, {Pool},
  {Popa}, {Popa}, {Popp}, {Porciani}, {Porth}, {Potter}, {Poulain},
  {Pourtsidou}, {Pozzetti}, {Prandoni}, {Pratt}, {Prezelus}, {Prieto}, {Pugno},
  {Quai}, {Quilley}, {Racca}, {Raccanelli}, {R{\'a}cz}, {Radinovi{\'c}},
  {Radovich}, {Ragagnin}, {Ragnit}, {Raison}, {Ramos-Chernenko}, {Ranc},
  {Raylet}, {Rebolo}, {Refregier}, {Reimberg}, {Reiprich}, {Renk}, {Renzi},
  {Retre}, {Revaz}, {Reyl{\'e}}, {Reynolds}, {Rhodes}, {Ricci}, {Ricci},
  {Riccio}, {Ricken}, {Rissanen}, {Risso}, {Rix}, {Robin}, {Rocca-Volmerange},
  {Rocci}, {Rodenhuis}, {Rodighiero}, {Rodriguez Monroy}, {Rollins},
  {Romanello}, {Roman}, {Romelli}, {Romero-Gomez}, {Roncarelli}, {Rosati},
  {Rosset}, {Rossetti}, {Roster}, {Rottgering}, {Rozas-Fern{\'a}ndez}, {Ruane},
  {Rubino-Martin}, {Rudolph}, {Ruppin}, {Rusholme}, {Sacquegna},
  {S{\'a}ez-Casares}, {Saga}, {Saglia}, {Sahl{\'e}n}, {Saifollahi}, {Sakr},
  {Salvalaggio}, {Salvaterra}, {Salvati}, {Salvato}, {Salvignol},
  {S{\'a}nchez}, {Sanchez}, {Sanders}, {Sapone}, {Saponara}, {Sarpa}, {Sarron},
  {Sartori}, {Sassolas}, {Sauniere}, {Sauvage}, {Sawicki}, {Scaramella},
  {Scarlata}, {Scharr{\'e}}, {Schaye}, {Schewtschenko}, {Schindler},
  {Schinnerer}, {Schirmer}, {Schmidt}, {Schmidt}, {Schmidt}, {Schneider},
  {Schneider}, {Schneider}, {Sch{\"o}neberg}, {Schrabback}, {Schultheis},
  {Schulz}, {Schwartz}, {Sciotti}, {Scodeggio}, {Scognamiglio}, {Scott},
  {Scottez}, {Secroun}, {Sefusatti}, {Seidel}, {Seiffert}, {Sellentin},
  {Selwood}, {Semboloni}, {Sereno}, {Serjeant}, {Serrano}, {Shankar},
  {Sharples}, {Short}, {Shulevski}, {Shuntov}, {Sias}, {Sikkema}, {Silvestri},
  {Simon}, {Sirignano}, {Sirri}, {Skottfelt}, {Slezak}, {Sluse}, {Smith},
  {Smith}, {Smith}, {Smit}, {Soldano}, {Solheim}, {Sorce}, {Sorrenti},
  {Soubrie}, {Spinoglio}, {Spurio Mancini}, {Stadel}, {Stagnaro}, {Stanco},
  {Stanford}, {Starck}, {Stassi}, {Steinwagner}, {Stern}, {Stone}, {Strada},
  {Strafella}, {Stramaccioni}, {Surace}, {Sureau}, {Suyu}, {Swindells},
  {Szafraniec}, {Szapudi}, {Taamoli}, {Talia}, {Tallada-Cresp{\'\i}},
  {Tanidis}, {Tao}, {Tarr{\'\i}o}, {Tavagnacco}, {Taylor}, {Taylor}, {Taylor},
  {Teixeira}, {Tenti}, {Teodoro Idiago}, {Teplitz}, {Tereno}, {Tessore},
  {Testa}, {Testera}, {Tewes}, {Teyssier}, {Theret}, {Thizy}, {Thomas}, {Toba},
  {Toft}, {Toledo-Moreo}, {Tolstoy}, {Tommasi}, {Torbaniuk}, {Torradeflot},
  {Tortora}, {Tosi}, {Tosti}, {Trifoglio}, {Troja}, {Trombetti}, {Tronconi},
  {Tsedrik}, {Tsyganov}, {Tucci}, {Tutusaus}, {Uhlemann}, {Ulivi}, {Urbano},
  {Vacher}, {Vaillon}, {Valdes}, {Valentijn}, {Valenziano}, {Valieri},
  {Valiviita}, {Van den Broeck}, {Vassallo}, {Vavrek}, {Venemans}, {Venhola},
  {Ventura}, {Verdoes Kleijn}, {Vergani}, {Verma}, {Vernizzi}, {Veropalumbo},
  {Verza}, {Vescovi}, {Vibert}, {Viel}, {Vielzeuf}, {Viglione}, {Viitanen},
  {Villaescusa-Navarro}, {Vinciguerra}, {Visticot}, {Voggel}, {von
  Wietersheim-Kramsta}, {Vriend}, {Wachter}, {Walmsley}, {Walth}, {Walton},
  {Walton}, {Wander}, {Wang}, {Wang}, {Weaver}, {Weller}, {Whalen}, {Wiesmann},
  {Wilde}, {Williams}, {Winther}, {Wittje}, {Wong}, {Wright}, {Yankelevich},
  {Yeung}, {Youles}, {Yung}, {Zacchei}, {Zalesky}, {Zamorani}, {Zamorano
  Vitorelli}, {Zanoni Marc}, {Zennaro}, {Zerbi}, {Zinchenko}, {Zoubian},
  {Zucca}, \& {Zumalacarregui}}]{euclid}
{Euclid Collaboration}, {Mellier}, Y., {Abdurro'uf}, {et~al.} 2024, arXiv
  e-prints, arXiv:2405.13491, \dodoi{10.48550/arXiv.2405.13491}

\bibitem[{{Garling} {et~al.}(2024){Garling}, {Peter}, {Spekkens}, {Sand},
  {Hargis}, {Crnojevi{\'c}}, \& {Carlin}}]{Garling24}
{Garling}, C.~T., {Peter}, A. H.~G., {Spekkens}, K., {et~al.} 2024, \mnras,
  528, 365, \dodoi{10.1093/mnras/stae014}

\bibitem[{{Gatto} {et~al.}(2024){Gatto}, {Bellazzini}, {Tortora}, {Ripepi},
  {Dall'Ora}, {Cignoni}, {Kuijken}, {Hildebrandt}, {Zhang}, {de Jong},
  {Napolitano}, \& {Smith}}]{Gatto24}
{Gatto}, M., {Bellazzini}, M., {Tortora}, C., {et~al.} 2024, \aap, 681, L13,
  \dodoi{10.1051/0004-6361/202348554}

\bibitem[{{Geha} {et~al.}(2012){Geha}, {Blanton}, {Yan}, \& {Tinker}}]{Geha12}
{Geha}, M., {Blanton}, M.~R., {Yan}, R., \& {Tinker}, J.~L. 2012, \apj, 757,
  85, \dodoi{10.1088/0004-637X/757/1/85}

\bibitem[{{Giovanelli} {et~al.}(2013){Giovanelli}, {Haynes}, {Adams}, {Cannon},
  {Rhode}, {Salzer}, {Skillman}, {Bernstein-Cooper}, \&
  {McQuinn}}]{Giovanelli13}
{Giovanelli}, R., {Haynes}, M.~P., {Adams}, E. A.~K., {et~al.} 2013, \aj, 146,
  15, \dodoi{10.1088/0004-6256/146/1/15}

\bibitem[{{Hargis} {et~al.}(2020){Hargis}, {Albers}, {Crnojevi{\'c}}, {Sand},
  {Weisz}, {Carlin}, {Spekkens}, {Willman}, {Peter}, {Grillmair}, \&
  {Dolphin}}]{Hargis20}
{Hargis}, J.~R., {Albers}, S., {Crnojevi{\'c}}, D., {et~al.} 2020, \apj, 888,
  31, \dodoi{10.3847/1538-4357/ab58d2}

\bibitem[{{Homma} {et~al.}(2024){Homma}, {Chiba}, {Komiyama}, {Tanaka},
  {Okamoto}, {Tanaka}, {Ishigaki}, {Hayashi}, {Arimoto}, {Lupton}, {Strauss},
  {Miyazaki}, {Wang}, \& {Murayama}}]{Homma24}
{Homma}, D., {Chiba}, M., {Komiyama}, Y., {et~al.} 2024, \pasj, 76, 733,
  \dodoi{10.1093/pasj/psae044}

\bibitem[{{Hook} {et~al.}(2004){Hook}, {J{\o}rgensen}, {Allington-Smith},
  {Davies}, {Metcalfe}, {Murowinski}, \& {Crampton}}]{GMOS}
{Hook}, I.~M., {J{\o}rgensen}, I., {Allington-Smith}, J.~R., {et~al.} 2004,
  \pasp, 116, 425, \dodoi{10.1086/383624}

\bibitem[{{Hopkins} {et~al.}(2012){Hopkins}, {Quataert}, \&
  {Murray}}]{Hopkins12}
{Hopkins}, P.~F., {Quataert}, E., \& {Murray}, N. 2012, \mnras, 421, 3522,
  \dodoi{10.1111/j.1365-2966.2012.20593.x}

\bibitem[{{Iglesias-P{\'a}ramo} {et~al.}(2006){Iglesias-P{\'a}ramo}, {Buat},
  {Takeuchi}, {Xu}, {Boissier}, {Boselli}, {Burgarella}, {Madore}, {Gil de
  Paz}, {Bianchi}, {Barlow}, {Byun}, {Donas}, {Forster}, {Friedman}, {Heckman},
  {Jelinski}, {Lee}, {Malina}, {Martin}, {Milliard}, {Morrissey}, {Neff},
  {Rich}, {Schiminovich}, {Seibert}, {Siegmund}, {Small}, {Szalay}, {Welsh}, \&
  {Wyder}}]{Iglesias-Paramo+2006}
{Iglesias-P{\'a}ramo}, J., {Buat}, V., {Takeuchi}, T.~T., {et~al.} 2006, \apjs,
  164, 38, \dodoi{10.1086/502628}

\bibitem[{{Into} \& {Portinari}(2013)}]{Into13}
{Into}, T., \& {Portinari}, L. 2013, \mnras, 430, 2715,
  \dodoi{10.1093/mnras/stt071}

\bibitem[{{Ivezi{\'c}} {et~al.}(2019){Ivezi{\'c}}, {Kahn}, {Tyson}, {Abel},
  {Acosta}, {Allsman}, {Alonso}, {AlSayyad}, {Anderson}, {Andrew}, {Angel},
  {Angeli}, {Ansari}, {Antilogus}, {Araujo}, {Armstrong}, {Arndt}, {Astier},
  {Aubourg}, {Auza}, {Axelrod}, {Bard}, {Barr}, {Barrau}, {Bartlett}, {Bauer},
  {Bauman}, {Baumont}, {Bechtol}, {Bechtol}, {Becker}, {Becla}, {Beldica},
  {Bellavia}, {Bianco}, {Biswas}, {Blanc}, {Blazek}, {Blandford}, {Bloom},
  {Bogart}, {Bond}, {Booth}, {Borgland}, {Borne}, {Bosch}, {Boutigny},
  {Brackett}, {Bradshaw}, {Brandt}, {Brown}, {Bullock}, {Burchat}, {Burke},
  {Cagnoli}, {Calabrese}, {Callahan}, {Callen}, {Carlin}, {Carlson},
  {Chandrasekharan}, {Charles-Emerson}, {Chesley}, {Cheu}, {Chiang}, {Chiang},
  {Chirino}, {Chow}, {Ciardi}, {Claver}, {Cohen-Tanugi}, {Cockrum}, {Coles},
  {Connolly}, {Cook}, {Cooray}, {Covey}, {Cribbs}, {Cui}, {Cutri}, {Daly},
  {Daniel}, {Daruich}, {Daubard}, {Daues}, {Dawson}, {Delgado}, {Dellapenna},
  {de Peyster}, {de Val-Borro}, {Digel}, {Doherty}, {Dubois},
  {Dubois-Felsmann}, {Durech}, {Economou}, {Eifler}, {Eracleous}, {Emmons},
  {Fausti Neto}, {Ferguson}, {Figueroa}, {Fisher-Levine}, {Focke}, {Foss},
  {Frank}, {Freemon}, {Gangler}, {Gawiser}, {Geary}, {Gee}, {Geha}, {Gessner},
  {Gibson}, {Gilmore}, {Glanzman}, {Glick}, {Goldina}, {Goldstein}, {Goodenow},
  {Graham}, {Gressler}, {Gris}, {Guy}, {Guyonnet}, {Haller}, {Harris},
  {Hascall}, {Haupt}, {Hernandez}, {Herrmann}, {Hileman}, {Hoblitt}, {Hodgson},
  {Hogan}, {Howard}, {Huang}, {Huffer}, {Ingraham}, {Innes}, {Jacoby}, {Jain},
  {Jammes}, {Jee}, {Jenness}, {Jernigan}, {Jevremovi{\'c}}, {Johns}, {Johnson},
  {Johnson}, {Jones}, {Juramy-Gilles}, {Juri{\'c}}, {Kalirai}, {Kallivayalil},
  {Kalmbach}, {Kantor}, {Karst}, {Kasliwal}, {Kelly}, {Kessler}, {Kinnison},
  {Kirkby}, {Knox}, {Kotov}, {Krabbendam}, {Krughoff}, {Kub{\'a}nek},
  {Kuczewski}, {Kulkarni}, {Ku}, {Kurita}, {Lage}, {Lambert}, {Lange},
  {Langton}, {Le Guillou}, {Levine}, {Liang}, {Lim}, {Lintott}, {Long},
  {Lopez}, {Lotz}, {Lupton}, {Lust}, {MacArthur}, {Mahabal}, {Mandelbaum},
  {Markiewicz}, {Marsh}, {Marshall}, {Marshall}, {May}, {McKercher}, {McQueen},
  {Meyers}, {Migliore}, {Miller}, {Mills}, {Miraval}, {Moeyens}, {Moolekamp},
  {Monet}, {Moniez}, {Monkewitz}, {Montgomery}, {Morrison}, {Mueller},
  {Muller}, {Mu{\~n}oz Arancibia}, {Neill}, {Newbry}, {Nief}, {Nomerotski},
  {Nordby}, {O'Connor}, {Oliver}, {Olivier}, {Olsen}, {O'Mullane}, {Ortiz},
  {Osier}, {Owen}, {Pain}, {Palecek}, {Parejko}, {Parsons}, {Pease},
  {Peterson}, {Peterson}, {Petravick}, {Libby Petrick}, {Petry},
  {Pierfederici}, {Pietrowicz}, {Pike}, {Pinto}, {Plante}, {Plate}, {Plutchak},
  {Price}, {Prouza}, {Radeka}, {Rajagopal}, {Rasmussen}, {Regnault}, {Reil},
  {Reiss}, {Reuter}, {Ridgway}, {Riot}, {Ritz}, {Robinson}, {Roby}, {Roodman},
  {Rosing}, {Roucelle}, {Rumore}, {Russo}, {Saha}, {Sassolas}, {Schalk},
  {Schellart}, {Schindler}, {Schmidt}, {Schneider}, {Schneider}, {Schoening},
  {Schumacher}, {Schwamb}, {Sebag}, {Selvy}, {Sembroski}, {Seppala}, {Serio},
  {Serrano}, {Shaw}, {Shipsey}, {Sick}, {Silvestri}, {Slater}, {Smith},
  {Smith}, {Sobhani}, {Soldahl}, {Storrie-Lombardi}, {Stover}, {Strauss},
  {Street}, {Stubbs}, {Sullivan}, {Sweeney}, {Swinbank}, {Szalay}, {Takacs},
  {Tether}, {Thaler}, {Thayer}, {Thomas}, {Thornton}, {Thukral}, {Tice},
  {Trilling}, {Turri}, {Van Berg}, {Vanden Berk}, {Vetter}, {Virieux},
  {Vucina}, {Wahl}, {Walkowicz}, {Walsh}, {Walter}, {Wang}, {Wang}, {Warner},
  {Wiecha}, {Willman}, {Winters}, {Wittman}, {Wolff}, {Wood-Vasey}, {Wu},
  {Xin}, {Yoachim}, \& {Zhan}}]{lsst}
{Ivezi{\'c}}, {\v{Z}}., {Kahn}, S.~M., {Tyson}, J.~A., {et~al.} 2019, \apj,
  873, 111, \dodoi{10.3847/1538-4357/ab042c}

\bibitem[{{Jeon} {et~al.}(2017){Jeon}, {Besla}, \& {Bromm}}]{Jeon17}
{Jeon}, M., {Besla}, G., \& {Bromm}, V. 2017, \apj, 848, 85,
  \dodoi{10.3847/1538-4357/aa8c80}

\bibitem[{{Jones} {et~al.}(2023){Jones}, {Mutlu-Pakdil}, {Sand}, {Donnerstein},
  {Crnojevi{\'c}}, {Bennet}, {Fielder}, {Karunakaran}, {Spekkens}, {Strader},
  {Urquhart}, \& {Zaritsky}}]{Jones23}
{Jones}, M.~G., {Mutlu-Pakdil}, B., {Sand}, D.~J., {et~al.} 2023, \apjl, 957,
  L5, \dodoi{10.3847/2041-8213/ad0130}

\bibitem[{{Jones} {et~al.}(2024){Jones}, {Sand}, {Mutlu-Pakdil}, {Fielder},
  {Crnojevi{\'c}}, {Bennet}, {Spekkens}, {Donnerstein}, {Doliva-Dolinsky},
  {Karunakaran}, {Strader}, \& {Zaritsky}}]{Jones24}
{Jones}, M.~G., {Sand}, D.~J., {Mutlu-Pakdil}, B., {et~al.} 2024, \apjl, 971,
  L37, \dodoi{10.3847/2041-8213/ad676e}

\bibitem[{{Jordi} {et~al.}(2006){Jordi}, {Grebel}, \& {Ammon}}]{Jordi06}
{Jordi}, K., {Grebel}, E.~K., \& {Ammon}, K. 2006, \aap, 460, 339,
  \dodoi{10.1051/0004-6361:20066082}

\bibitem[{{Kalberla} \& {Haud}(2015)}]{Kalberla15}
{Kalberla}, P.~M.~W., \& {Haud}, U. 2015, \aap, 578, A78,
  \dodoi{10.1051/0004-6361/201525859}

\bibitem[{{Kalberla} {et~al.}(2010){Kalberla}, {McClure-Griffiths}, {Pisano},
  {Calabretta}, {Ford}, {Lockman}, {Staveley-Smith}, {Kerp}, {Winkel},
  {Murphy}, \& {Newton-McGee}}]{Kalberla10}
{Kalberla}, P.~M.~W., {McClure-Griffiths}, N.~M., {Pisano}, D.~J., {et~al.}
  2010, \aap, 521, A17, \dodoi{10.1051/0004-6361/200913979}

\bibitem[{{Karachentsev} \& {Kaisina}(2019)}]{Kara19}
{Karachentsev}, I.~D., \& {Kaisina}, E.~I. 2019, Astrophysical Bulletin, 74,
  111, \dodoi{10.1134/S1990341319020019}

\bibitem[{{Karachentsev} {et~al.}(2004){Karachentsev}, {Karachentseva},
  {Huchtmeier}, \& {Makarov}}]{Kara04}
{Karachentsev}, I.~D., {Karachentseva}, V.~E., {Huchtmeier}, W.~K., \&
  {Makarov}, D.~I. 2004, \aj, 127, 2031, \dodoi{10.1086/382905}

\bibitem[{{Karunakaran} {et~al.}(2021){Karunakaran}, {Spekkens}, {Oman},
  {Simpson}, {Fattahi}, {Sand}, {Bennet}, {Crnojevi{\'c}}, {Frenk},
  {G{\'o}mez}, {Grand}, {Jones}, {Marinacci}, {Mutlu-Pakdil}, {Navarro}, \&
  {Zaritsky}}]{Karunakaran21}
{Karunakaran}, A., {Spekkens}, K., {Oman}, K.~A., {et~al.} 2021, \apjl, 916,
  L19, \dodoi{10.3847/2041-8213/ac0e3a}

\bibitem[{{Koposov} {et~al.}(2018){Koposov}, {Walker}, {Belokurov}, {Casey},
  {Geringer-Sameth}, {Mackey}, {Da Costa}, {Erkal}, {Jethwa}, {Mateo},
  {Olszewski}, \& {Bailey}}]{Koposov18}
{Koposov}, S.~E., {Walker}, M.~G., {Belokurov}, V., {et~al.} 2018, \mnras, 479,
  5343, \dodoi{10.1093/mnras/sty1772}

\bibitem[{{Labrie} {et~al.}(2023{\natexlab{a}}){Labrie}, {Simpson}, {Cardenes},
  {Turner}, {Soraisam}, {Quint}, {Oberdorf}, {Placco}, {Berke}, {Smirnova},
  {Conseil}, {Vacca}, \& {Thomas-Osip}}]{dragonsRNAAS_2023}
{Labrie}, K., {Simpson}, C., {Cardenes}, R., {et~al.} 2023{\natexlab{a}},
  Research Notes of the American Astronomical Society, 7, 214,
  \dodoi{10.3847/2515-5172/ad0044}

\bibitem[{{Labrie} {et~al.}(2023{\natexlab{b}}){Labrie}, {Simpson}, {Turner},
  {Quint}, {Conseil}, {Oberdorf}, {Soraisam}, {Placco}, {Smirnova}, {Berke}, \&
  {Vacca}}]{dragons3.1.0}
{Labrie}, K., {Simpson}, C., {Turner}, J., {et~al.} 2023{\natexlab{b}},
  {DRAGONS}, 3.1.0, Zenodo,  Zenodo, \dodoi{10.5281/zenodo.7776065}

\bibitem[{{Landsman}(1993)}]{IDLforever}
{Landsman}, W.~B. 1993, in Astronomical Society of the Pacific Conference
  Series, Vol.~52, Astronomical Data Analysis Software and Systems II, ed.
  R.~J. {Hanisch}, R.~J.~V. {Brissenden}, \& J.~{Barnes}, 246

\bibitem[{{Lang} {et~al.}(2010){Lang}, {Hogg}, {Mierle}, {Blanton}, \&
  {Roweis}}]{astrometry}
{Lang}, D., {Hogg}, D.~W., {Mierle}, K., {Blanton}, M., \& {Roweis}, S. 2010,
  \aj, 139, 1782, \dodoi{10.1088/0004-6256/139/5/1782}

\bibitem[{{Lee} {et~al.}(2011){Lee}, {Gil de Paz}, {Kennicutt}, {Bothwell},
  {Dalcanton}, {Jos{\'e} G. Funes S.}, {Johnson}, {Sakai}, {Skillman},
  {Tremonti}, \& {van Zee}}]{Lee11}
{Lee}, J.~C., {Gil de Paz}, A., {Kennicutt}, Robert~C., J., {et~al.} 2011,
  \apjs, 192, 6, \dodoi{10.1088/0067-0049/192/1/6}

\bibitem[{{Li} {et~al.}(2024){Li}, {Greene}, {Carlsten}, \& {Danieli}}]{Li24}
{Li}, J., {Greene}, J.~E., {Carlsten}, S.~G., \& {Danieli}, S. 2024, arXiv
  e-prints, arXiv:2406.00101, \dodoi{10.48550/arXiv.2406.00101}

\bibitem[{{Mac Low} \& {Ferrara}(1999)}]{Maclow99}
{Mac Low}, M.-M., \& {Ferrara}, A. 1999, \apj, 513, 142, \dodoi{10.1086/306832}

\bibitem[{{Manwadkar} \& {Kravtsov}(2022)}]{Manwadkar22}
{Manwadkar}, V., \& {Kravtsov}, A.~V. 2022, \mnras, 516, 3944,
  \dodoi{10.1093/mnras/stac2452}

\bibitem[{{Mao} {et~al.}(2021){Mao}, {Geha}, {Wechsler}, {Weiner}, {Tollerud},
  {Nadler}, \& {Kallivayalil}}]{Mao21}
{Mao}, Y.-Y., {Geha}, M., {Wechsler}, R.~H., {et~al.} 2021, \apj, 907, 85,
  \dodoi{10.3847/1538-4357/abce58}

\bibitem[{{Mao} {et~al.}(2024){Mao}, {Geha}, {Wechsler}, {Asali}, {Wang},
  {Kado-Fong}, {Kallivayalil}, {Nadler}, {Tollerud}, {Weiner}, {de los Reyes},
  \& {Wu}}]{Mao24}
---. 2024, arXiv e-prints, arXiv:2404.14498, \dodoi{10.48550/arXiv.2404.14498}

\bibitem[{{Martin} {et~al.}(2005){Martin}, {Fanson}, {Schiminovich},
  {Morrissey}, {Friedman}, {Barlow}, {Conrow}, {Grange}, {Jelinsky},
  {Milliard}, {Siegmund}, {Bianchi}, {Byun}, {Donas}, {Forster}, {Heckman},
  {Lee}, {Madore}, {Malina}, {Neff}, {Rich}, {Small}, {Surber}, {Szalay},
  {Welsh}, \& {Wyder}}]{galex}
{Martin}, D.~C., {Fanson}, J., {Schiminovich}, D., {et~al.} 2005, \apjl, 619,
  L1, \dodoi{10.1086/426387}

\bibitem[{{Martin} {et~al.}(2008){Martin}, {de Jong}, \& {Rix}}]{Martin08}
{Martin}, N.~F., {de Jong}, J. T.~A., \& {Rix}, H.-W. 2008, \apj, 684, 1075,
  \dodoi{10.1086/590336}

\bibitem[{{Martin} {et~al.}(2013){Martin}, {Ibata}, {McConnachie}, {Mackey},
  {Ferguson}, {Irwin}, {Lewis}, \& {Fardal}}]{Martin13}
{Martin}, N.~F., {Ibata}, R.~A., {McConnachie}, A.~W., {et~al.} 2013, \apj,
  776, 80, \dodoi{10.1088/0004-637X/776/2/80}

\bibitem[{{McClure-Griffiths} {et~al.}(2009){McClure-Griffiths}, {Pisano},
  {Calabretta}, {Ford}, {Lockman}, {Staveley-Smith}, {Kalberla}, {Bailin},
  {Dedes}, {Janowiecki}, {Gibson}, {Murphy}, {Nakanishi}, \&
  {Newton-McGee}}]{McClure-Griffiths09}
{McClure-Griffiths}, N.~M., {Pisano}, D.~J., {Calabretta}, M.~R., {et~al.}
  2009, \apjs, 181, 398, \dodoi{10.1088/0067-0049/181/2/398}

\bibitem[{{McConnachie}(2012)}]{Mcconachie12}
{McConnachie}, A.~W. 2012, \aj, 144, 4, \dodoi{10.1088/0004-6256/144/1/4}

\bibitem[{{McConnachie} \& {Irwin}(2006)}]{McConnachie06}
{McConnachie}, A.~W., \& {Irwin}, M.~J. 2006, \mnras, 365, 1263,
  \dodoi{10.1111/j.1365-2966.2005.09806.x}

\bibitem[{{McNanna} {et~al.}(2024){McNanna}, {Bechtol}, {Mau}, {Nadler},
  {Medoff}, {Drlica-Wagner}, {Cerny}, {Crnojevi{\'c}}, {Mutlu-Pakd{\i}l},
  {Vivas}, {Pace}, {Carlin}, {Collins}, {Ferguson}, {Mart{\'\i}nez-Delgado},
  {Mart{\'\i}nez-V{\'a}zquez}, {Noel}, {Riley}, {Sand}, {Smercina}, {Tollerud},
  {Wechsler}, {Abbott}, {Aguena}, {Alves}, {Bacon}, {Bom}, {Brooks}, {Burke},
  {Carballo-Bello}, {Carnero Rosell}, {Carretero}, {da Costa}, {Davis}, {de
  Vicente}, {Diehl}, {Doel}, {Ferrero}, {Frieman}, {Giannini}, {Gruen},
  {Gutierrez}, {Gruendl}, {Hinton}, {Hollowood}, {Honscheid}, {James}, {Kuehn},
  {Marshall}, {Mena-Fern{\'a}ndez}, {Miquel}, {Pereira}, {Pieres},
  {Malag{\'o}n}, {Sakowska}, {Sanchez}, {Sanchez Cid}, {Santiago},
  {Sevilla-Noarbe}, {Smith}, {Stringfellow}, {Suchyta}, {Swanson}, {Tarle},
  {Weaverdyck}, {Wiseman}, {DES Collaboration}, \& {DELVE
  Collaboration}}]{McNanna24}
{McNanna}, M., {Bechtol}, K., {Mau}, S., {et~al.} 2024, \apj, 961, 126,
  \dodoi{10.3847/1538-4357/ad07d0}

\bibitem[{{McQuinn} {et~al.}(2023){McQuinn}, {Mao}, {Buckley}, {Shih}, {Cohen},
  \& {Dolphin}}]{McQuinn23}
{McQuinn}, K. B.~W., {Mao}, Y.-Y., {Buckley}, M.~R., {et~al.} 2023, \apj, 944,
  14, \dodoi{10.3847/1538-4357/acaec9}

\bibitem[{{McQuinn} {et~al.}(2024){McQuinn}, {Mao}, {Tollerud}, {Cohen},
  {Shih}, {Buckley}, \& {Dolphin}}]{McQuinn24}
{McQuinn}, K. B.~W., {Mao}, Y.-Y., {Tollerud}, E.~J., {et~al.} 2024, \apj, 967,
  161, \dodoi{10.3847/1538-4357/ad429b}

\bibitem[{{McQuinn} {et~al.}(2014){McQuinn}, {Cannon}, {Dolphin}, {Skillman},
  {Salzer}, {Haynes}, {Adams}, {Cave}, {Elson}, {Giovanelli}, {Ott}, \&
  {Saintonge}}]{McQuinn14}
{McQuinn}, K. B.~W., {Cannon}, J.~M., {Dolphin}, A.~E., {et~al.} 2014, \apj,
  785, 3, \dodoi{10.1088/0004-637X/785/1/3}

\bibitem[{{McQuinn} {et~al.}(2015){McQuinn}, {Skillman}, {Dolphin}, {Cannon},
  {Salzer}, {Rhode}, {Adams}, {Berg}, {Giovanelli}, {Girardi}, \&
  {Haynes}}]{McQuinn15}
{McQuinn}, K. B.~W., {Skillman}, E.~D., {Dolphin}, A., {et~al.} 2015, \apj,
  812, 158, \dodoi{10.1088/0004-637X/812/2/158}

\bibitem[{{McQuinn} {et~al.}(2021){McQuinn}, {Telidevara}, {Fuson}, {Adams},
  {Cannon}, {Skillman}, {Dolphin}, {Haynes}, {Rhode}, {Salzer}, {Giovanelli},
  \& {Gordon}}]{McQuinn21}
{McQuinn}, K. B.~W., {Telidevara}, A.~K., {Fuson}, J., {et~al.} 2021, \apj,
  918, 23, \dodoi{10.3847/1538-4357/ac03ae}

\bibitem[{{Moster} {et~al.}(2010){Moster}, {Somerville}, {Maulbetsch}, {van den
  Bosch}, {Macci{\`o}}, {Naab}, \& {Oser}}]{Moster10}
{Moster}, B.~P., {Somerville}, R.~S., {Maulbetsch}, C., {et~al.} 2010, \apj,
  710, 903, \dodoi{10.1088/0004-637X/710/2/903}

\bibitem[{{Mu{\~n}oz} {et~al.}(2018){Mu{\~n}oz}, {C{\^o}t{\'e}}, {Santana},
  {Geha}, {Simon}, {Oyarz{\'u}n}, {Stetson}, \& {Djorgovski}}]{Munoz18}
{Mu{\~n}oz}, R.~R., {C{\^o}t{\'e}}, P., {Santana}, F.~A., {et~al.} 2018, \apj,
  860, 66, \dodoi{10.3847/1538-4357/aac16b}

\bibitem[{{Mu{\~n}oz-Mateos} {et~al.}(2015){Mu{\~n}oz-Mateos}, {Sheth},
  {Regan}, {Kim}, {Laine}, {Erroz-Ferrer}, {Gil de Paz}, {Comeron}, {Hinz},
  {Laurikainen}, {Salo}, {Athanassoula}, {Bosma}, {Bouquin}, {Schinnerer},
  {Ho}, {Zaritsky}, {Gadotti}, {Madore}, {Holwerda}, {Men{\'e}ndez-Delmestre},
  {Knapen}, {Meidt}, {Querejeta}, {Mizusawa}, {Seibert}, {Laine}, \&
  {Courtois}}]{Munoz15}
{Mu{\~n}oz-Mateos}, J.~C., {Sheth}, K., {Regan}, M., {et~al.} 2015, \apjs, 219,
  3, \dodoi{10.1088/0067-0049/219/1/3}

\bibitem[{{Mutlu-Pakdil} {et~al.}(2018){Mutlu-Pakdil}, {Sand}, {Carlin},
  {Spekkens}, {Caldwell}, {Crnojevi{\'c}}, {Hughes}, {Willman}, \&
  {Zaritsky}}]{Mutlu18}
{Mutlu-Pakdil}, B., {Sand}, D.~J., {Carlin}, J.~L., {et~al.} 2018, \apj, 863,
  25, \dodoi{10.3847/1538-4357/aacd0e}

\bibitem[{{Mutlu-Pakdil} {et~al.}(2021){Mutlu-Pakdil}, {Sand}, {Crnojevi{\'c}},
  {Drlica-Wagner}, {Caldwell}, {Guhathakurta}, {Seth}, {Simon}, {Strader}, \&
  {Toloba}}]{Mutlu_sims}
{Mutlu-Pakdil}, B., {Sand}, D.~J., {Crnojevi{\'c}}, D., {et~al.} 2021, \apj,
  918, 88, \dodoi{10.3847/1538-4357/ac0db8}

\bibitem[{{Mutlu-Pakdil} {et~al.}(2024){Mutlu-Pakdil}, {Sand}, {Crnojevi{\'c}},
  {Bennet}, {Jones}, {Spekkens}, {Karunakaran}, {Zaritsky}, {Caldwell},
  {Fielder}, {Guhathakurta}, {Seth}, {Simon}, {Strader}, \& {Toloba}}]{Mutlu24}
---. 2024, arXiv e-prints, arXiv:2401.14457, \dodoi{10.48550/arXiv.2401.14457}

\bibitem[{{Putman} {et~al.}(2021){Putman}, {Zheng}, {Price-Whelan}, {Grcevich},
  {Johnson}, {Tollerud}, \& {Peek}}]{Putnam21}
{Putman}, M.~E., {Zheng}, Y., {Price-Whelan}, A.~M., {et~al.} 2021, \apj, 913,
  53, \dodoi{10.3847/1538-4357/abe391}

\bibitem[{{Rey} {et~al.}(2020){Rey}, {Pontzen}, {Agertz}, {Orkney}, {Read}, \&
  {Rosdahl}}]{Rey20}
{Rey}, M.~P., {Pontzen}, A., {Agertz}, O., {et~al.} 2020, \mnras, 497, 1508,
  \dodoi{10.1093/mnras/staa1640}

\bibitem[{{Rey} {et~al.}(2022){Rey}, {Pontzen}, {Agertz}, {Orkney}, {Read},
  {Saintonge}, {Kim}, \& {Das}}]{Rey22}
---. 2022, \mnras, 511, 5672, \dodoi{10.1093/mnras/stac502}

\bibitem[{{Richstein} {et~al.}(2024){Richstein}, {Kallivayalil}, {Simon},
  {Garling}, {Wetzel}, {Warfield}, {van der Marel}, {Jeon}, {Rose}, {Torrey},
  {Engelhardt}, {Besla}, {Choi}, {Geha}, {Guhathakurta}, {Kirby}, {Patel},
  {Sacchi}, \& {Sohn}}]{Richstein24}
{Richstein}, H., {Kallivayalil}, N., {Simon}, J.~D., {et~al.} 2024, \apj, 967,
  72, \dodoi{10.3847/1538-4357/ad393c}

\bibitem[{{Ricotti} \& {Gnedin}(2005)}]{Ricotti05}
{Ricotti}, M., \& {Gnedin}, N.~Y. 2005, \apj, 629, 259, \dodoi{10.1086/431415}

\bibitem[{{Rodriguez Wimberly} {et~al.}(2019){Rodriguez Wimberly}, {Cooper},
  {Fillingham}, {Boylan-Kolchin}, {Bullock}, \&
  {Garrison-Kimmel}}]{Rodriguez19}
{Rodriguez Wimberly}, M.~K., {Cooper}, M.~C., {Fillingham}, S.~P., {et~al.}
  2019, \mnras, 483, 4031, \dodoi{10.1093/mnras/sty3357}

\bibitem[{{Sacchi} {et~al.}(2021){Sacchi}, {Richstein}, {Kallivayalil}, {van
  der Marel}, {Libralato}, {Zivick}, {Besla}, {Brown}, {Choi}, {Deason},
  {Fritz}, {Geha}, {Guhathakurta}, {Jeon}, {Kirby}, {Majewski}, {Patel},
  {Simon}, {Tony Sohn}, {Tollerud}, \& {Wetzel}}]{Sacchi21}
{Sacchi}, E., {Richstein}, H., {Kallivayalil}, N., {et~al.} 2021, \apjl, 920,
  L19, \dodoi{10.3847/2041-8213/ac2aa3}

\bibitem[{{Sales} {et~al.}(2022){Sales}, {Wetzel}, \& {Fattahi}}]{Sales22}
{Sales}, L.~V., {Wetzel}, A., \& {Fattahi}, A. 2022, Nature Astronomy, 6, 897,
  \dodoi{10.1038/s41550-022-01689-w}

\bibitem[{{Samuel} {et~al.}(2020){Samuel}, {Wetzel}, {Chapman}, {Tollerud},
  {Hopkins}, {Boylan-Kolchin}, {Bailin}, \& {Faucher-Gigu{\`e}re}}]{Samuel20}
{Samuel}, J., {Wetzel}, A., {Chapman}, S., {et~al.} 2020, arXiv e-prints,
  arXiv:2010.08571.
\newblock \doarXiv{2010.08571}

\bibitem[{{Sand} {et~al.}(2009){Sand}, {Olszewski}, {Willman}, {Zaritsky},
  {Seth}, {Harris}, {Piatek}, \& {Saha}}]{Sand09}
{Sand}, D.~J., {Olszewski}, E.~W., {Willman}, B., {et~al.} 2009, \apj, 704,
  898, \dodoi{10.1088/0004-637X/704/2/898}

\bibitem[{{Sand} {et~al.}(2015{\natexlab{a}}){Sand}, {Spekkens},
  {Crnojevi{\'c}}, {Hargis}, {Willman}, {Strader}, \& {Grillmair}}]{Sand15}
{Sand}, D.~J., {Spekkens}, K., {Crnojevi{\'c}}, D., {et~al.}
  2015{\natexlab{a}}, \apjl, 812, L13, \dodoi{10.1088/2041-8205/812/1/L13}

\bibitem[{{Sand} {et~al.}(2012){Sand}, {Strader}, {Willman}, {Zaritsky},
  {McLeod}, {Caldwell}, {Seth}, \& {Olszewski}}]{Sand12}
{Sand}, D.~J., {Strader}, J., {Willman}, B., {et~al.} 2012, \apj, 756, 79,
  \dodoi{10.1088/0004-637X/756/1/79}

\bibitem[{{Sand} {et~al.}(2014){Sand}, {Crnojevi{\'c}}, {Strader}, {Toloba},
  {Simon}, {Caldwell}, {Guhathakurta}, {McLeod}, \& {Seth}}]{Sand14}
{Sand}, D.~J., {Crnojevi{\'c}}, D., {Strader}, J., {et~al.} 2014, \apjl, 793,
  L7, \dodoi{10.1088/2041-8205/793/1/L7}

\bibitem[{{Sand} {et~al.}(2015{\natexlab{b}}){Sand}, {Crnojevi{\'c}}, {Bennet},
  {Willman}, {Hargis}, {Strader}, {Olszewski}, {Tollerud}, {Simon}, {Caldwell},
  {Guhathakurta}, {James}, {Koposov}, {McLeod}, {Morrell}, {Peacock},
  {Salinas}, {Seth}, {Stark}, \& {Toloba}}]{Sand_UCHVC}
{Sand}, D.~J., {Crnojevi{\'c}}, D., {Bennet}, P., {et~al.} 2015{\natexlab{b}},
  \apj, 806, 95, \dodoi{10.1088/0004-637X/806/1/95}

\bibitem[{{Sand} {et~al.}(2022){Sand}, {Mutlu-Pakdil}, {Jones}, {Karunakaran},
  {Wang}, {Yang}, {Chiti}, {Bennet}, {Crnojevi{\'c}}, \& {Spekkens}}]{Sand22}
{Sand}, D.~J., {Mutlu-Pakdil}, B., {Jones}, M.~G., {et~al.} 2022, \apjl, 935,
  L17, \dodoi{10.3847/2041-8213/ac85ee}

\bibitem[{{Santos-Santos} {et~al.}(2023){Santos-Santos}, {Navarro}, \&
  {McConnachie}}]{Santos23}
{Santos-Santos}, I. M.~E., {Navarro}, J.~F., \& {McConnachie}, A. 2023, \mnras,
  520, 55, \dodoi{10.1093/mnras/stad085}

\bibitem[{{Saviane} {et~al.}(1996){Saviane}, {Held}, \& {Piotto}}]{Savian96}
{Saviane}, I., {Held}, E.~V., \& {Piotto}, G. 1996, \aap, 315, 40.
\newblock \doarXiv{astro-ph/9601165}

\bibitem[{{Sawala} {et~al.}(2016){Sawala}, {Frenk}, {Fattahi}, {Navarro},
  {Bower}, {Crain}, {Dalla Vecchia}, {Furlong}, {Helly}, {Jenkins}, {Oman},
  {Schaller}, {Schaye}, {Theuns}, {Trayford}, \& {White}}]{Sawala16}
{Sawala}, T., {Frenk}, C.~S., {Fattahi}, A., {et~al.} 2016, \mnras, 457, 1931,
  \dodoi{10.1093/mnras/stw145}

\bibitem[{{Schlafly} \& {Finkbeiner}(2011)}]{Schlafly11}
{Schlafly}, E.~F., \& {Finkbeiner}, D.~P. 2011, \apj, 737, 103,
  \dodoi{10.1088/0004-637X/737/2/103}

\bibitem[{{Simon}(2019)}]{Simon19}
{Simon}, J.~D. 2019, \araa, 57, 375,
  \dodoi{10.1146/annurev-astro-091918-104453}

\bibitem[{{Simon} {et~al.}(2021){Simon}, {Brown}, {Drlica-Wagner}, {Li},
  {Avila}, {Bechtol}, {Clementini}, {Crnojevi{\'c}}, {Garofalo}, {Geha},
  {Sand}, {Strader}, \& {Willman}}]{Simon21}
{Simon}, J.~D., {Brown}, T.~M., {Drlica-Wagner}, A., {et~al.} 2021, \apj, 908,
  18, \dodoi{10.3847/1538-4357/abd31b}

\bibitem[{{Slater} \& {Bell}(2014)}]{Slater14}
{Slater}, C.~T., \& {Bell}, E.~F. 2014, \apj, 792, 141,
  \dodoi{10.1088/0004-637X/792/2/141}

\bibitem[{{Smercina} {et~al.}(2018){Smercina}, {Bell}, {Price}, {D'Souza},
  {Slater}, {Bailin}, {Monachesi}, \& {Nidever}}]{Smercina18}
{Smercina}, A., {Bell}, E.~F., {Price}, P.~A., {et~al.} 2018, \apj, 863, 152,
  \dodoi{10.3847/1538-4357/aad2d6}

\bibitem[{{Smith} {et~al.}(2023){Smith}, {Jensen}, {Roediger}, {Sestito},
  {Hayes}, {McConnachie}, {Cuillandre}, {Gwyn}, {Magnier}, {Chambers},
  {Hammer}, {Hudson}, {Martin}, {Navarro}, \& {Scott}}]{Smith23}
{Smith}, S. E.~T., {Jensen}, J., {Roediger}, J., {et~al.} 2023, \aj, 166, 76,
  \dodoi{10.3847/1538-3881/acdd77}

\bibitem[{{Smith} {et~al.}(2024){Smith}, {Cerny}, {Hayes}, {Sestito}, {Jensen},
  {McConnachie}, {Geha}, {Navarro}, {Li}, {Cuillandre}, {Errani}, {Chambers},
  {Gwyn}, {Hammer}, {Hudson}, {Magnier}, \& {Martin}}]{Smith24}
{Smith}, S. E.~T., {Cerny}, W., {Hayes}, C.~R., {et~al.} 2024, \apj, 961, 92,
  \dodoi{10.3847/1538-4357/ad0d9f}

\bibitem[{{Stetson}(1987)}]{Stetson87}
{Stetson}, P.~B. 1987, \pasp, 99, 191, \dodoi{10.1086/131977}

\bibitem[{{Stetson}(1994)}]{Stetson94}
---. 1994, \pasp, 106, 250, \dodoi{10.1086/133378}

\bibitem[{{Teyssier} {et~al.}(2012){Teyssier}, {Johnston}, \&
  {Kuhlen}}]{Teyssier12}
{Teyssier}, M., {Johnston}, K.~V., \& {Kuhlen}, M. 2012, \mnras, 426, 1808,
  \dodoi{10.1111/j.1365-2966.2012.21793.x}

\bibitem[{{The Astropy Collaboration} {et~al.}(2018){The Astropy
  Collaboration}, {Price-Whelan}, {Sip{\H o}cz}, {G{\"u}nther}, {Lim},
  {Crawford}, {Conseil}, {Shupe}, {Craig}, {Dencheva}, {Ginsburg},
  {VanderPlas}, {Bradley}, {P{\'e}rez-Su{\'a}rez}, {de Val-Borro}, {Aldcroft},
  {Cruz}, {Robitaille}, {Tollerud}, {Ardelean}, {Babej}, {Bachetti}, {Bakanov},
  {Bamford}, {Barentsen}, {Barmby}, {Baumbach}, {Berry}, {Biscani}, {Boquien},
  {Bostroem}, {Bouma}, {Brammer}, {Bray}, {Breytenbach}, {Buddelmeijer},
  {Burke}, {Calderone}, {Cano Rodr{\'{\i}}guez}, {Cara}, {Cardoso},
  {Cheedella}, {Copin}, {Crichton}, {D{\'A}vella}, {Deil}, {Depagne},
  {Dietrich}, {Donath}, {Droettboom}, {Earl}, {Erben}, {Fabbro}, {Ferreira},
  {Finethy}, {Fox}, {Garrison}, {Gibbons}, {Goldstein}, {Gommers}, {Greco},
  {Greenfield}, {Groener}, {Grollier}, {Hagen}, {Hirst}, {Homeier}, {Horton},
  {Hosseinzadeh}, {Hu}, {Hunkeler}, {Ivezi{\'c}}, {Jain}, {Jenness}, {Kanarek},
  {Kendrew}, {Kern}, {Kerzendorf}, {Khvalko}, {King}, {Kirkby}, {Kulkarni},
  {Kumar}, {Lee}, {Lenz}, {Littlefair}, {Ma}, {Macleod}, {Mastropietro},
  {McCully}, {Montagnac}, {Morris}, {Mueller}, {Mumford}, {Muna}, {Murphy},
  {Nelson}, {Nguyen}, {Ninan}, {N{\"o}the}, {Ogaz}, {Oh}, {Parejko}, {Parley},
  {Pascual}, {Patil}, {Patil}, {Plunkett}, {Prochaska}, {Rastogi}, {Reddy
  Janga}, {Sabater}, {Sakurikar}, {Seifert}, {Sherbert}, {Sherwood-Taylor},
  {Shih}, {Sick}, {Silbiger}, {Singanamalla}, {Singer}, {Sladen}, {Sooley},
  {Sornarajah}, {Streicher}, {Teuben}, {Thomas}, {Tremblay}, {Turner},
  {Terr{\'o}n}, {van Kerkwijk}, {de la Vega}, {Watkins}, {Weaver}, {Whitmore},
  {Woillez}, \& {Zabalza}}]{astropy}
{The Astropy Collaboration}, {Price-Whelan}, A.~M., {Sip{\H o}cz}, B.~M.,
  {et~al.} 2018, ArXiv e-prints.
\newblock \doarXiv{1801.02634}

\bibitem[{{Tollerud} {et~al.}(2015){Tollerud}, {Geha}, {Grcevich}, {Putman}, \&
  {Stern}}]{Tollerud15}
{Tollerud}, E.~J., {Geha}, M.~C., {Grcevich}, J., {Putman}, M.~E., \& {Stern},
  D. 2015, \apjl, 798, L21, \dodoi{10.1088/2041-8205/798/1/L21}

\bibitem[{{Tollerud} \& {Peek}(2018)}]{Tollerud18}
{Tollerud}, E.~J., \& {Peek}, J.~E.~G. 2018, \apj, 857, 45,
  \dodoi{10.3847/1538-4357/aab3e4}

\bibitem[{{Torrealba} {et~al.}(2018){Torrealba}, {Belokurov}, {Koposov},
  {Bechtol}, {Drlica-Wagner}, {Olsen}, {Vivas}, {Yanny}, {Jethwa}, {Walker},
  {Li}, {Allam}, {Conn}, {Gallart}, {Gruendl}, {James}, {Johnson}, {Kuehn},
  {Kuropatkin}, {Martin}, {Martinez-Delgado}, {Nidever}, {No{\"e}l}, {Simon},
  {Stringfellow}, \& {Tucker}}]{Torrealba18}
{Torrealba}, G., {Belokurov}, V., {Koposov}, S.~E., {et~al.} 2018, \mnras, 475,
  5085, \dodoi{10.1093/mnras/sty170}

\bibitem[{{van der Marel} {et~al.}(2002){van der Marel}, {Alves}, {Hardy}, \&
  {Suntzeff}}]{vdMarel02}
{van der Marel}, R.~P., {Alves}, D.~R., {Hardy}, E., \& {Suntzeff}, N.~B. 2002,
  \aj, 124, 2639, \dodoi{10.1086/343775}

\bibitem[{{Walsh} {et~al.}(2008){Walsh}, {Willman}, {Sand}, {Harris}, {Seth},
  {Zaritsky}, \& {Jerjen}}]{Walsh08}
{Walsh}, S.~M., {Willman}, B., {Sand}, D., {et~al.} 2008, \apj, 688, 245,
  \dodoi{10.1086/592076}

\bibitem[{{Weerasooriya} {et~al.}(2023){Weerasooriya}, {Bovill}, {Benson},
  {Musick}, \& {Ricotti}}]{Weerasooriya23}
{Weerasooriya}, S., {Bovill}, M.~S., {Benson}, A., {Musick}, A.~M., \&
  {Ricotti}, M. 2023, \apj, 948, 87, \dodoi{10.3847/1538-4357/acc32b}

\bibitem[{{Weisz} \& {Boylan-Kolchin}(2019)}]{Weisz19}
{Weisz}, D., \& {Boylan-Kolchin}, M. 2019, \baas, 51, 1.
\newblock \doarXiv{1901.07571}

\bibitem[{{Weisz} {et~al.}(2014{\natexlab{a}}){Weisz}, {Dolphin}, {Skillman},
  {Holtzman}, {Gilbert}, {Dalcanton}, \& {Williams}}]{Weisz14}
{Weisz}, D.~R., {Dolphin}, A.~E., {Skillman}, E.~D., {et~al.}
  2014{\natexlab{a}}, \apj, 789, 147, \dodoi{10.1088/0004-637X/789/2/147}

\bibitem[{{Weisz} {et~al.}(2011){Weisz}, {Dalcanton}, {Williams}, {Gilbert},
  {Skillman}, {Seth}, {Dolphin}, {McQuinn}, {Gogarten}, {Holtzman}, {Rosema},
  {Cole}, {Karachentsev}, \& {Zaritsky}}]{Weisz11}
{Weisz}, D.~R., {Dalcanton}, J.~J., {Williams}, B.~F., {et~al.} 2011, \apj,
  739, 5, \dodoi{10.1088/0004-637X/739/1/5}

\bibitem[{{Weisz} {et~al.}(2014{\natexlab{b}}){Weisz}, {Skillman}, {Hidalgo},
  {Monelli}, {Dolphin}, {McConnachie}, {Bernard}, {Gallart}, {Aparicio},
  {Boylan-Kolchin}, {Cassisi}, {Cole}, {Ferguson}, {Irwin}, {Martin}, {Mayer},
  {McQuinn}, {Navarro}, \& {Stetson}}]{Weisz14M31}
{Weisz}, D.~R., {Skillman}, E.~D., {Hidalgo}, S.~L., {et~al.}
  2014{\natexlab{b}}, \apj, 789, 24, \dodoi{10.1088/0004-637X/789/1/24}

\bibitem[{{Wetzel} {et~al.}(2016){Wetzel}, {Hopkins}, {Kim},
  {Faucher-Gigu{\`e}re}, {Kere{\v{s}}}, \& {Quataert}}]{Wetzel16}
{Wetzel}, A.~R., {Hopkins}, P.~F., {Kim}, J.-h., {et~al.} 2016, \apjl, 827,
  L23, \dodoi{10.3847/2041-8205/827/2/L23}

\end{thebibliography}
\bibliographystyle{aasjournal}

\end{document}